\documentclass[]{aa}  
\usepackage{graphicx}
\usepackage{txfonts}

\usepackage{epsf}
\usepackage[authoryear]{natbib}
\usepackage{astro_bib_macro}
\usepackage{amssymb}
\bibpunct{(}{)}{;}{a}{}{,}
\usepackage[centerlast,it,small]{subfigure}
\usepackage{aalongtable}
\usepackage{supertabular}
\usepackage{lscape}
\usepackage{rotating}
\newcommand{\kms}{km\,s$^{-1}$}
\newcommand{\fuse}{\emph{FUSE}}

\newcommand{\iue}{\emph{IUE}}

\newcommand{\bpic}{$\beta$ Pictoris}

\newcommand{\lyalpha}{\mbox{Ly$\alpha$}}

\newcommand{\lb}{$\lambda$}

\newcommand{\vsini}{$v$\,sin\,$i$}   
\newcommand{\vrad}{v$_{\rm rad}$}

\newcommand{\teff}{T$_{\rm eff}$}

\newcommand{\owens}{{\sc Owens}}
\newcommand{\Av}{A$_{\rm v}$}

\begin{document}
\title{Molecular hydrogen in the circumstellar environments of Herbig
  Ae/Be stars probed by FUSE}
   
   \author{C. Martin-Za\"{\i}di\inst{1}       
          \and
           M. Deleuil\inst{2}
	  \and
	   J. Le Bourlot\inst{3}
           \and
           J.-C. Bouret\inst{2}
          \and
	   A. Roberge\inst{4}
          \and
	   C.P. Dullemond\inst{5}
           \and
           L. Testi\inst{6}
          \and
	   P.D. Feldman\inst{7}
           \and
	   A. Lecavelier des Etangs\inst{8}
          \and
	   A. Vidal-Madjar\inst{8}}
          
   \offprints{C. Martin-Za\"{\i}di}

   \institute{Laboratoire d’Astrophysique de Grenoble, CNRS,
     Université Joseph-Fourier, UMR5571, Grenoble, France  \\
 \email{claire.martin-zaidi@obs.ujf-grenoble.fr}
    \and
    Laboratoire d'Astrophysique de Marseille, BP
    8, Les trois Lucs, 13376 Marseille Cedex 12, France
    \and 
    LUTH, Observatoire de Paris, Universit\'e Paris 7, 92195 Meudon
    Cedex, France
    \and 
     NASA Goddard Space Flight Center, Greenbelt, MD 20771, USA 
     \and
     Max-Planck-Institut f\"ur Astronomie, K\"onigstuhl 17, D-69117
     Heidelberg, Germany
    \and 
    Osservatorio Astrofisico di Arcetri, Largo E. Fermi 5, 50125
    Firenze, Italy
    \and 
     Department of Physics and Astronomy, JHU, Baltimore, MD 21218, USA
    \and 
    Institut d'Astrophysique de Paris, CNRS, 98bis Boulevard Arago,
    75014 Paris, France  }

   \date{Received ... / Accepted ...}

\abstract
{Molecular hydrogen (H$_2$) gas is the most abundant molecule in the
  circumstellar (CS) environments of young stars. It is thus a key
  element in our understanding of the evolution of pre-main sequence
  stars and their environments towards the main sequence.}
{At the present time, little is known about the gas as compared to the
  dust in the environments of young stars. We thus observed molecular
  hydrogen around a sample of pre-main sequence stars in order to
  better characterize their gaseous CS environments.}
{The \fuse\ ({\it Far Ultraviolet Spectroscopic Explorer}) spectral
  range offers access to electronic transitions of H$_2$.  We analyzed
  the \fuse\ spectra of a sample of Herbig Ae/Be stars (HAeBes)
  covering a broad spectral range (from F4 to B2), including the
  main-sequence A5 star $\beta$ Pictoris. To better diagnose the
  origin of the detected molecular gas and its excitation conditions,
  we used a model of a photodissociation region.}
{Our analysis demonstrates that the excitation of H$_2$ is clearly
  different around most of the HAeBes compared to the interstellar
  medium. Moreover, the characteristics of H$_2$ around Herbig Ae and
  Be stars give evidence for different excitation mechanisms. For the
  most massive stars of our sample (B8 to B2 type), the excitation
  diagrams are reproduced well by a model of photodissociation regions
  (PDR). Our results favor an interpretation in terms of large CS
  envelopes, remnants of the molecular clouds in which the stars were
  formed. On the other hand, the group of Ae stars (later than B9
  type) known to possess disks is more inhomogeneous. In most cases,
  when CS H$_2$ is detected, the lines of sight do not pass through
  the disks. The excitation conditions of H$_2$ around Ae stars cannot
  be reproduced by PDR models and correspond to warm and/or hot
  excited media very close to the stars. In addition, no clear
  correlation has been found between the ages of the stars and the
  amount of circumstellar H$_2$. Our results suggest structural
  differences between Herbig Ae and Be star environments. Herbig Be
  stars do evolve faster than Ae stars, and consequently, most Herbig
  Be stars are younger than Ae ones at the time we observe them. It is
  thus more likely to find remnants of their parent cloud around
  them.}
   {}

\keywords{stars: circumstellar matter -- stars: formation -- stars:
pre-main sequence -- ISM: molecules }

\titlerunning{H$_2$ in the CS Environment of HAeBes}
\authorrunning{C. Martin-Za\"{\i}di et al.}

\maketitle

\section{Introduction}

A long standing problem in modern astrophysics is how stars and
planets form from their parent molecular clouds. It is generally
accepted that the collapse of an interstellar molecular cloud core to
form a protostar naturally produces a disk-shaped nebula in which
planets might form. In the past twenty years, both planets and disks
have been observed around nearby stars. Some young stars with disks,
e.g., the T Tauri star GM Aur \citep{Rice03}, are suspected of
harboring young planets, and recently a massive planet orbiting TW
Hya, a young star surrounded by a well-studied disk, has been
discovered by means radial velocity techniques \citep{Setiawan08}.

Disks have been directly imaged around T Tauri stars \citep[e.g., HK
Tau,][]{Stapelfeldt98}, while direct or coronagraphic imaging of
Herbig Ae/Be star disks has been performed successfully only in a few
cases \citep[e.g.][]{Pantin00, GRADY01, Augereau04, Lagage06}. Most
evidence of disks around Herbig Ae/Be stars is indirect: blueshifted
absorption lines, polarization, mass estimates from millimeter
observations, extinction measurements, and simple disk-model fitting
to infrared spectral energy distributions (IR SEDs) \citep{Bastien90,
  HILL92, CORCORAN97}. Numerous observations have revealed elongated
structures around several HAeBes, with velocity gradients along the
major axes characteristic of gas in Keplerian rotation
\citep{Mannings97, Mannings00}. In addition, the SEDs of some HAeBes
show dips near 10$\mu$m that cannot be explained by normal
dust-removal processes, such as the Poynting-Robertson drag and
radiation pressure from the star. It instead suggests that the dust
structure is evolving because of the disk breaking up. One possibility
is that this is caused by planet formation \citep{Bouwman03}. In this
context, HAeBes are possible precursors of $\beta$ Pictoris and
Vega-type stars, whose CS debris disks are believed to host planetary
bodies. This raises the possibility that the environment around the
HAeBes truly represents a very early phase of planet formation. The
physical parameters of HAeBes disks can thus be used to constrain the
duration of the protoplanetary phase and the age at which the
signatures of planets become visible in CS disks.

On the other hand, mid-IR observations at high spatial resolution have
revealed that the emission observed from HAeBes is generally not
confined to optically thick disks but instead comes from more complex
environments such as remnant envelopes \citep{POLOMSKI02}. On the
basis of millimeter interferometric measurements, \citet{NATTA00}
conclude that CS envelopes are much more common around Herbig Be stars
(HBes) than around Herbig Ae/B9 stars (HAes). These conclusions have
been confirmed by near-IR speckel interferometry
\citep{LEINERT01}. This indicates structural differences between
Herbig Ae stars and Herbig Be stars. These results are fully
compatible with the faster evolution of the more massive HBes.
Indeed, due to their stronger radiation fields, the CS environments
around HBes evolve faster than around HAes, which translates
especially into a rapid depletion of the CS disk material.
Consequently, the average HBe stars are younger than HAes at the time
we observe them; therefore, it is more likely to find larger amounts
of remnant CS material around Herbig Be stars, since less time is
available to dissipate it. In this scenario, it is expected that the
CS material is mostly concentrated at the outer edges of the close CS
environment and distributed in more or less spherically envelopes.

Molecular hydrogen, from which giant planets are primarily believed to
form, is the most abundant molecule in the CS environments of young
stars. The detection of H$_2$ provides the most direct information
about the gaseous content of the CS environments of HAeBes, setting
limits on the timescales for the dissipation of CS matter and possible
planet building. Tentative detections of pure rotational H$_2$ lines
at infrared wavelengths towards a few Herbig stars and T Tauri stars
have been previously obtained with ISO-SWS \citep{Thi01}. But
ground-based observations have shown that contamination from
surrounding cloud material can be important and that ISO could not
have detected disk gas \citep{Richter02, Sako05}. The H$_2$ mid-IR
emission lines have been clearly detected in the disks of only two
Herbig stars, namely AB Aur \citep{Bitner07} and HD~97048
\citep{klr07a}. These studies show that the gas has not dissipated in
the inner parts of these disks with ages of about 3 Myrs and that
peculiar H$_2$ excitation conditions are present in the disks. The
electronic transitions of H$_2$, which are much stronger than the
rotational ones, occur in the far-ultraviolet (FUV) spectral range, a
domain covered by \fuse\ ({\it Far Ultraviolet Spectrocopic
  Explorer}). Previous \fuse\ observations of selected young stars
harboring CS disks \citep{LECAV01, LECAV03, ROBERGE01} have
demonstrated that such a study can make significant contributions
clarifying issues raised in previous work.

We present the results of our systematic analysis of the \fuse\
spectra of a large group of Herbig HAeBe stars with a variety of
spectral types and Vega-type stars, including $\beta$ Pictoris. This
paper is organized as follows. In Sect.~\ref{data} we present our
sample stars and recall the methods used for data reduction and
analysis, which have been presented in previous papers. The results
obtained from the H$_2$ analysis are given in Sect.~\ref{H2}. The
detailed analysis and interpretation of our results for each group of
stars are presented in Sects.~\ref{disks} and \ref{HBe}. We explore
possible evolutionary trends in our results in Sect.~\ref{evol}.  Our
conclusions are discussed in Sect.~\ref{discu}.


\begin{table*}
\begin{center}
  \caption{Log of the observations.}
\begin{tabular}{lccclccc}
\noalign{\medskip}\hline
\hline
Star        &  $\alpha$   & $\delta$     & \fuse\       &  Principal              & Total observing  & \fuse    & Published   \\
            & (2000)      & (2000)       & Program      &  Investigator $^{{\mathrm(a)}}$ &  time (ks)       & aperture & Data $^{{\mathrm(b)}}$       \\
\hline
$\beta$ Pictoris & 05 47 17.09 & -51 03 59.45 & P219         & A. Vidal-Madjar          &  24.78         & LWRS      & (1)   \\ 
            &             &              & Q119         & M. Deleuil               &  34.50         & LWRS      &     \\ 
            &             &              & C132         & J.-C. Bouret             &  47.79         & LWRS      &      \\ 
HD~135344    & 15 15 48.44 & -37 09 16.03 & Q306         & A. Lecavelier des Etangs &  8.44          & LWRS      &     \\ 
HD~100453    & 11 33 05.57 & -54 19 28.54 & C126         & archives       &  11.79         & LWRS      &      \\ 
HD~36112     & 05 30 27.53 & +25 19 57.08 & Q319         & M. Deleuil               &  7.10          & LWRS      &     \\     
HD~104237    & 12 00 05.08 & -78 11 34.56 & P163         & archives   &  18.97         & LWRS      &     \\     
            &             &              & P263         & archives   &  20.84         & LWRS      &     \\   
HD~163296    & 17 56 21.29 & -21 57 21.88 & P219         & A. Vidal-Madjar          &  15.92         & LWRS      & (2)    \\
            &             &              & Q219         & A. Lecavelier des Etangs &  16.16         & LWRS      &     \\ 
NX~PUP      & 07 19 28.26 & -44 35 11.28 & Z906         & Observatory Program      &  17.80         & LWRS      &     \\   
AB~Aur      & 04 55 45.84 & +30 33 04.29 & P119         & A. Vidal-Madjar          &  13.66         & LWRS      & (3)  \\
            &             &              & P219         & A. Vidal-Madjar          &  15.55         & LWRS      &    \\ 
HD~141569    & 15 49 57.75 & -03 55 16.36 & Q319         & M. Deleuil               &  6.79          & LWRS      & (4)   \\
HD~100546    & 11 33 25.44 & -70 11 41.24 & P119         & A. Vidal-Madjar          &  10.35         & LWRS      & (5)  \\
            &             &              & P219         & A. Vidal-Madjar          &  11.94         & LWRS      &      \\ 
HD~109573    & 12 36 01.03 & -39 52 10.22 & B091         & archives       &  12.38         & LWRS      &    \\   
HD~176386    & 19 01 38.93 & -36 53 26.55 & P119         & A. Vidal-Madjar          &  14.98         & LWRS      &      \\
            &             &              & P219         & A. Vidal-Madjar          &  27.39         & LWRS      &     \\ 
HD~250550    & 06 01 59.99 & +16 30 56.73 & B038         & C. Catala                &  9.06          & LWRS      & (6)   \\
HD~85567     & 09 50 28.54 & -60 58 02.96 & Z906         & Observatory Program      &  7.06          & LWRS      &    \\
HD~259431    & 06 33 05.19 & +10 19 19.98 & B038         & C. Catala                &  16.68         & LWRS      & (6)    \\
HD~38087     & 05 43 00.57 & -02 18 45.37 & A063         & archives      &  31.12         & LWRS      &     \\
            &             &              & P116         & archives     &  4.00          & MDRS      &    \\
HD~76534     & 08 55 08.71 & -43 27 59.86 & B038         & C. Catala                &  5.71          & LWRS      & (7)   \\
\hline
\end{tabular}
\begin{list}{}{}
\item $^{{\mathrm(a)}}$ Most of the data were obtained by our Team
  (through guaranteed time and guest investigator observations). Some
  of the data are from the \fuse\ archives or from observatory
  programs dedicated to instrument testing.
\item $^{{\mathrm(b)}}$: data already published.
\item {\it References}: (1) \cite{LECAV01}; (2) \cite{D05}; (3)
  \cite{ROBERGE01}; (4) \cite{klr05}; (5) \cite{D04}; (6)
  \cite{JCB03}; (7) \cite{klr04}.
\end{list}
\label{tab_obs}
\end{center}
\end{table*}


\section{The sample stars}
\label{data}

To perform this study, we have compiled a set of observations from
different programs we carried out over the past six years,
supplemented by some data from the archives (Table~\ref{tab_obs}).
This results in a sample of 18 HAeBes with spectral types from F4 to
B2 and one main-sequence star, $\beta$ Pictoris, prototype of the
Vega-type stars. Two stars, HD~141569A and HD~109573, belong to the
so-called transitional class and are passing from the pre-main
sequence Herbig star stage to the zero-age main sequence (ZAMS)
stars. Both are known to possess CS disks \citep{Jura91, Augereau99}
with ages close to about 10 Myrs (Table~\ref{tab_sample}).


\begin{figure}[!htbp] 
  \includegraphics[width=9cm]{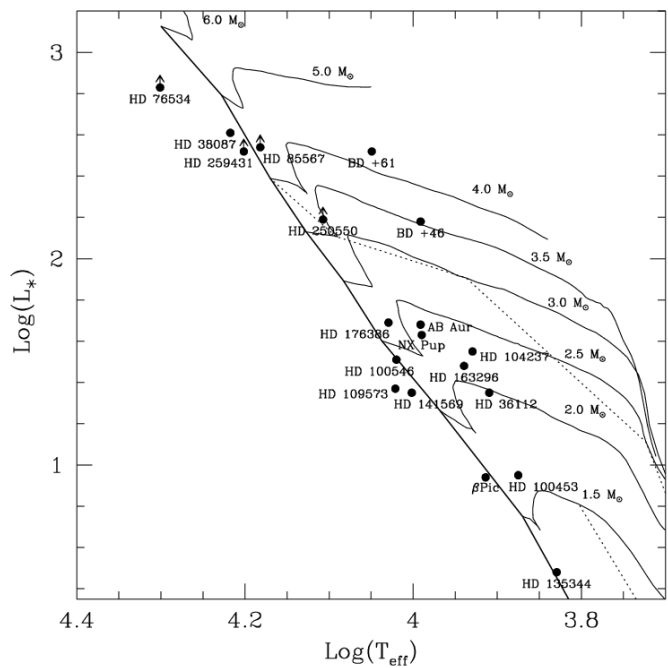}
  \caption{The stars of our sample are plotted on this HR diagram. The
    evolutionary tracks between the pre-main sequence phase and the
    ZAMS are shown for the Ae/B9 stars. For this plot, we used the
    \citet{PALLA93} pre-main sequence tracks and the interpolation
    routines written by \citet{TESTI98}. For the more massive Be
    stars, the pre-main sequence stage is less obvious due to the
    faster evolution of these stars, which makes them appear directly
    on the ZAMS. }
  \label{HR}
\end{figure}


Our sample also includes HD~135344. The spectral type of this star has
not been determined precisely in the literature, estimates ranging
from F4 to A0, but it has been classified as a Herbig star
\citep[e.g.][]{THE94, Malfait98, Thi01}.  In the rest of this paper,
we thus include this star in the sub sample called Herbig Ae/B9 stars.

For several stars, some early results of the \fuse\ spectral analysis
have already been published, and we will reiterate the results obtained
from these previous studies. The \fuse\ observations of 11 HAeBes are
presented here for the first time. Some stars have been observed twice
(e.g. AB~Aur) or even three times (e.g. \bpic, HD~104237) with a time
separation between the \fuse\ observations ranging from about one
month to more than one year. For the sake of consistency, we thus
re-did the spectral analysis of the entire sample and took advantage
of any new observations to improve the S/N of the spectra and the
H$_2$ spectral analysis.

The fundamental parameters of the stars are given in
Table~\ref{tab_sample}.  Most are from the literature, but in some
cases, we re-evaluated the effective temperatures of the stars
\citep[for details on the method, see][]{JCB03}. Using these values
and the \citet{PALLA93} pre-main sequence tracks, we calculated the
ages and masses using interpolation routines written by
\citet{TESTI98}. All the stars are plotted on the HR diagram presented
in Fig.~\ref{HR}, showing their evolutionary status.


\begin{table*}[!t]
\begin{center}
\vspace{1cm}
\caption{Physical parameters of the sample stars. }
\begin{tabular}{lclccrrcrcl}
\hline
\hline    
Name       & Spectral & T$_{eff}$               & E~(B-V) & \vsini               &  \vrad  &  Age~$^{\mathrm{(b)}}$ & Mass~$^{\mathrm{(b)}}$ & Distance   & Log L$_{\ast}$ & Ref. \\
           & Type     & (K)                     &         & (\kms)               & (\kms) & (Myr)                  & (M$_{\odot}$)          &   (pc)     & (L$_{\odot}$)  &     \\
\hline
$\beta$ Pictoris& A5V      & 8200                    & 0.02   & 140                   & +20    & 20                     &   1.75                 & 19.3       & 0.94           & {\tiny 1,2,3,4,5} \\
HD 135344  & F4V      & 6750                    & 0.15   & 80                    & -3     & 30                     &   1.3                  & 84         & 0.48           & {\tiny 1,6,7,8,9,11} \\
HD 100453  & A9       & 7500                    & 0.08   & 39                    & --     & 10                     &   1.7                  & 114        & 0.95           & {\tiny 1,7,8,10,11} \\
HD 36112   & A5       & 8120                    & 0.07   & 60                    & +17.6  & 5.0                    &   2.0                  & 200        & 1.35           & {\tiny 1,9,11,12,13} \\
HD 104237  & A4       & 8500                    & 0.10   & 10                    & +14    & 2.4                    &   2.4                  & 116        & 1.55           & {\tiny 1,7,11,12,14}  \\
HD 163296  & A1Ve     & 8700                    & 0.06   & 133                   & +4     & 4.7                    &   2.2                  & 122        & 1.48           & {\tiny 1,12,15,16} \\
NX PUP     & A0       & 9770                    & 0.19   & 120                   & --     & 2.8                    &   2.4                  & 450        & 1.63           & {\tiny 1,12,18,19,20}\\
AB Aur     & A0V      & 9800                    & 0.16   & 100                   & +21    &  2.6                   &   2.4                  & 144        & 1.68           & {\tiny 1,23,24,25}  \\
HD 141569  & B9V      & 10040                   & 0.13   & 236                   & -6.4   & $\geq$~5.0             &   2.3                  & 108        & 1.35           & {\tiny 1,12,28} \\
HD 100546  & B9V      & 10470                   & 0.08   & 55                    & +17    & $\geq$~9.0             &   2.4                  & 103        & 1.51           & {\tiny 1,11,12,29}  \\
HD 109573  & A0/B9    & 10500                   & 0.03   & 152                   & +9.4   & $\geq$~8.0             &   2.2                  & 67         & 1.37           & {\tiny 1,17,21,22} \\
HD 176386  & B9/B8    & 12000~$^{\mathrm{(a)}}$ & 0.19   & 220                   & +7.3   & 3.5                    &   2.7                  & 140        & 1.69           & {\tiny 1,30} \\
HD 250550  & B7       & 12800~$^{\mathrm{(a)}}$ & 0.22   & 110~$^{\mathrm{(a)}}$ & +31    & 1.0                    &   3.6                  & $\geq$~160 & 2.19           & {\tiny 1,12,23,32} \\
HD 85567   & B5V      & 15200~$^{\mathrm{(a)}}$ & 0.23   & 60~$^{\mathrm{(a)}}$  & 0/-5   & 1.0                    &   4.2                  & $\geq$~480 & 2.54           & {\tiny 1,12,33}  \\
HD 259431  & B5       & 15900~$^{\mathrm{(a)}}$ & 0.26   & 95~$^{\mathrm{(a)}}$  & +43    & $\geq$~1.0             &   4.4                  & 290-800    & 2.52           & {\tiny 1,12,23,32} \\
HD 38087   & B5V      & 16500~$^{\mathrm{(a)}}$ & 0.29   & 100~$^{\mathrm{(a)}}$ & +33    & $\geq$~1.0             &   4.6                  & 381        & 2.61           & {\tiny 1,14}~$^{\mathrm{(c)}}$\\
HD 76534   & B2       & 20000~$^{\mathrm{(a)}}$ & 0.32   & 110~$^{\mathrm{(a)}}$ & +17    & $\geq$~0.5             & $\geq$~5               & $\geq$~160 & 2.83           & {\tiny 1,12,23,34}  \\
\hline
\end{tabular}
\begin{list}{}{}
\item $^{\mathrm{(a)}}$: T$_{eff}$ and \vsini\ determined from our
  modeling \citep[see][for details of the method]{JCB03};
\item $^{\mathrm{(b)}}$: Ages and masses computed by us from
  \citet{PALLA93} tracks using the methods described in
  \citet{TESTI98};
\item $^{\mathrm{(c)}}$: Luminosity calculated using bolometric
  correction.
\item {\it References}: (1) SIMBAD data base; (2) \citet{Smith84}; (3)
  \citet{Barrado99}; (4) \citet{JCB02}; (5) \citet{crifo97}; (6)
  \citet{Malfait98}; (7) \citet{Meeus01}; (8) \citet{Dominik03}; (9)
  \citet{Dent05}; (10) \citet{Meeus02}; (11) \citet{Acke04c}; (12)
  \citet{VdA98}; (13) \citet{Beskrovnaya99}; (14) \citet{BERTOUT99};
  (15) \citet{Cidale01}; (16) \citet{Mora01}; (17) \citet{Torres03};
  (18) \citet{CORCORAN97}; (19) \citet{valenti00}; (20) \citet{BC95};
  (21) \citet{Royer02b}; (22) \citet{Gerbaldi99}; (23)
  \citet{FINK_JAN84}; (24) \citet{BC93}; (25) \citet{JCB98}; (26)
  \citet{Lada85}; (27) \citet{Hernandez04}; (28) \citet{Dunkin97};
  (29) \citet{Donati97}; (30) \citet{Siebenmorgen00}; (31)
  \citet{Millan-Gabet01}; (32) \citet{JCB03}; (33) \citet{MIRO01};
  (34) \citet{klr04}.
\end{list}
\label{tab_sample}
\end{center}
\end{table*}

\begin{table*}[]
\begin{center}
  \caption{Total column densities, radial velocities, intrinsic line
    widths $b$, and origin of the detected H$_2$ towards each star in
    the sample.}
\begin{tabular}{lccccccc}
\hline
\hline  
Name        &  log N(H$_2$) $^{\mathrm{(a)}}$             & \vrad\ (H$_2$) $^{\mathrm{(b)}}$     &  {\it b}             & Origin     & Ref.     & inclination          & Ref.  \\
            &                            &   (\kms)               &  (\kms)              & of H$_2$   &          & of the disk ({\degr}) $^{\mathrm{(c)}}$&   \\
\hline

$\beta$ Pictoris& $\leq$~17.41                &    --               &    --                &  --        & (1)      &  $\sim$0             & (6) \\

HD 135344  & $\leq$~17                   &    --            &    --                &   --       & (1)      & 79$\pm$2             & (7) \\
 
HD 100453  & $\leq$~16                   &   --             &    --                &    --      & (1)      & $\sim$65             & (8) \\

HD 36112   & $\leq$~17.85                &   ---             &   --                 &   --       & (1)      & 53-57                & (9) \\

HD 104237  & 18.68$_{-0.61}^{+0.39}$     & 1.4$_{-1.0}^{+0.9}$    & 5.3$_{-0.7}^{+0.6}$  & CS         & (1)      & 78                   & (10) \\

HD 163296  & 18.16$_{-0.40}^{+0.27}$     &   0$\pm$2              & 2.2$_{-0.7}^{+0.5}$  & CS         & (2), (1) & 60$\pm$5             &  (11)  \\

NX Pup     &  $\leq$~19.30               &    --               &   --                 &   --       &  (1)     &  ?                   &  \\

AB Aur     & 20.03$_{-0.19}^{+0.15}$     & 2$\pm$3               & 6.8$_{-2.2}^{+2.8}$  & IS/CS?     & (1),(4)  & 55-63                & (13)  \\


HD 141569  & 20.32$_{-0.22}^{+0.20}$     & 20                     & 5.3$_{-0.5}^{+0.3}$  &  IS        &  (1)     & 51$\pm$3             & (14) \\

HD 100546  & 16.46$_{-0.14}^{+0.24}$     & 0$\pm$2                & 3.4$_{-0.5}^{+0.8}$  & CS         & (2), (1) & 51$\pm$3             & (15)  \\

HD 109573  & $\leq$~15.40                &  --             &    --                &   --       & (1)      & 20.5$\pm$3           &  (12)    \\


HD 176386  & 20.80$_{-0.26}^{+0.18}$     & $\sim$~-7.0            &                      &  CS        & (1)      & n                    &  \\

HD 250550  & 19.26$_{-0.40}^{+0.17}$     & -1.0$_{-2.0}^{+1.7}$   & 5.6$_{-0.8}^{+0.7}$  & CS         & (3)      & n                    &     \\

HD 85567   & 19.33$_{-0.17}^{+0.20}$     & 4.5$_{-0.2}^{+0.3}$    & 6.0$_{-0.9}^{+0.7}$  & CS         & (1)      & n                    &    \\

HD 259431  & 20.64$_{-0.19}^{+0.11}$     & 13.0$_{-1.8}^{+2.2}$   & 4.5$_{-0.5}^{+0.3}$  & CS         & (3)      & n                    &      \\

HD 38087   & 20.43$_{-0.08}^{+0.15}$     & 2.36$_{-2.1}^{+1.2 }$  & 3.7$_{-0.2}^{+0.3}$  & CS         &  (1)     & n                    &       \\

HD 76534   & 20.64$_{-0.16}^{+0.16}$     & 0.0$_{-1.3}^{+0.3}$    & 5.0$_{-0.3}^{+0.4}$  & CS         & (5)      & n                    &       \\
\hline
\end{tabular}
\begin{list}{}{}
\item $^{\mathrm{(a)}}$ Upper limits on the column densities have been
  obtained with $b=1$~\kms.
\item $^{\mathrm{(b)}}$ Radial velocities are in the stars' rest
  frames.
\item $^{\mathrm{(c)}}$ The inclination angles for the disks are given
  with respect to the line of sight ($i=0$: edge-on). ``n'' indicates
  stars with no evidence of disk.
\item References in column 6 concern the results given in columns 2 to
  5, and references in column 8 are for the inclination angles of the
  disks. {\it References}: (1) this work; (2) \citet{LECAV03}; (3)
  \citet{JCB03}; (4) \citet{ROBERGE01}; (5) \citet{klr04}; (6)
  \citet{Smith84}; (7) \citet{Dent05}; (8) \citet{Dominik03}; (9)
  \citet{Eisner04}; (10) \citet{Grady04}; (11) \citet{GRADY00}; (12)
  \citet{Augereau99}; (13) \citet{Eisner03}; (14)
  \citet{Weinberger99}; (15) \citet{Augereau01}.
\end{list}
\label{results}
\end{center}
\end{table*}


All the stars were observed with the \fuse\ $30''\times30''$ LWRS
aperture at a resolution of about 15\,000. The \fuse\ observations
cover the wavelength range from 905 \AA\ to 1187 \AA. For the sake of
consistency, all the spectra were reprocessed with version 3.0.7 of
the \fuse\ calibration pipeline, CalFUSE \citep{Dixon07}. For each
star, the total exposure time was split into several sub-exposures
which have been co-aligned using a linear cross-correlation procedure
and added segment by segment. In some cases, the co-added spectra in
each detector channel were rebinned in wavelength to increase the S/N
ratio without degrading the resolution.


\section{Molecular hydrogen analysis } 
\label{H2}

\subsection{H$_2$ spectral analysis}
The \fuse\ spectral domain offers access to the Lyman and Werner
series electronic transitions of H$_2$. These electronic transitions
are between the rotational levels ($J$) of the ground vibrational
level ($v=0$) of the fundamental electronic level ($X$) and the
rotational levels of the vibrational levels of the first ($B$) and
second ($C$) excited electronic levels.  Under certain conditions,
spectral lines corresponding to transitions from the rotational levels
of higher vibrational levels ($v \geq 1$) of the fundamental
electronic level can also be observed \citep[e.g.][]{LECAV03, JCB03,
  Boisse05}.

For the hottest stars in our sample, typically stars whose \teff\
values are greater than 10\,000~K, the FUV stellar flux is high enough
to observe circumstellar and/or interstellar (CS/IS) lines in
absorption against the continuum. For the coolest stars, the lack of
sufficient flux in the continuum prevents easy detection of CS/IS
features. However, as highlighted by \cite{LECAV01}, stellar emission
lines, when present, can provide ``continuum'' flux for CS/IS
absorption features.  Most of the cool stars in our sample exhibit
broad emission in the \ion{O}{vi} \lb\lb1032-1038 resonance doublet
\cite[][]{D06}.  A number of H$_2$ transitions occur in the spectral
domain near 1034~\AA. In particular, the strongest transitions of
H$_2$, which arise from the $J=0$ to $J=2$ levels, fall in the same
wavelength range as the 1038~\AA\ emission line of the \ion{O}{vi}
doublet. This line is the weaker component of the \ion{O}{vi} doublet
and, as already shown by \cite{ROBERGE01}, the flux in the blue part
of the emission profile can be completely suppressed by strong H$_2$,
but also by \ion{C}{ii} absorption \citep{Roberge06}, making detection
of H$_2$ challenging. In that case, other lines corresponding to
transitions arising from upper J-levels should be observed,
especially, the spectral lines from transitions arising from the $J=3$
and $J=4$ rotational levels, which fall in the spectral domain of the
1032 \AA\ \ion{O}{vi} emission line.

We detected and measured H$_2$ absorption lines in the spectra of all
but 6 stars.  When H$_2$ is detected, we identified absorption lines
of H$_2$ arising from the rotational levels of the ground vibrational
state. Transitions arising from rotational levels as high as $J=5$ of
the first excited vibrational state ($v=1$) of the ground electronic
level are also detected in some cases. We performed molecular and
atomic absorption line analysis using the \owens\ profile fitting
procedure written by Dr. M. Lemoine \citep{LEMOIN02, HEBRARD02}, which
allowed us to simultaneously fit all lines of each species.  The
wavelengths and oscillator strengths of the H$_2$ lines were obtained
from \cite{ABGRA1} for the Lyman system and \cite{ABGRA2} for the
Werner system. The inverse of the total radiative lifetimes are
reported in \cite{ABGRA3}. For details on the line analysis procedure,
we refer the reader to our previous studies \citep[e.g.][]{JCB03,
  klr04, klr05}. We measured the column densities at each energy level
of H$_2$, the radial velocities, and the line widths. For most of the
stars, the H$_2$ mean projected velocity is very close to that of the
star, within the resolution of the \fuse\ spectra
($\sim$15~\kms). This is a clue that the detected gas is CS in
origin. Our results are given in Table~\ref{results} with 2-$\sigma$
error bars. Table~\ref{results2}, available as on-line material,
contains the following information. Columns 1 and 2 list the number
of the vibrational ($v$) and rotational ($J$) level, Column 3 gives
the statistical weight of each level, Column 4 gives the energy of
each level, and Columns 5 to 15 present the column densities of each
energy level of H$_2$, when detected, for all the targets stars.


\onltab{4}{
\begin{sidewaystable*}
 \begin{center}
  \caption{Column densities of the different energy levels of H$_2$
    when clearly observed in the \fuse\ spectra of the stars. ``g'' is
    the statistical weight of each level.}
\begin{tabular}{lccccccccccccccc}
\hline
\hline 
\noalign{\medskip}
   &    &     &           &  HD104237               & HD163296                &    AB Aur               & HD141569                & HD100546 & HD176386                & HD250550                & HD85567                 & HD259431                & HD38087                 & HD76534         \\
\noalign{\medskip}\hline
\noalign{\medskip}
$v$ & $J$ & $g$  & Energy    &            Log N        &        Log N            &        Log N            &        Log N            &        Log N  &        Log N        &        Log N            &        Log N            &        Log N            &        Log N  &        Log N    \\
   &    &     & level (K) &        (cm$^{-2}$)       &       (cm$^{-2}$)       &       (cm$^{-2}$)        &       (cm$^{-2}$)       &       (cm$^{-2}$)   &        (cm$^{-2}$)      &       (cm$^{-2}$)       &       (cm$^{-2}$)       &       (cm$^{-2}$)       &       (cm$^{-2}$)    &       (cm$^{-2}$)        \\
\noalign{\medskip}\hline
\noalign{\medskip}
 0 & 0  &  1  & 0.00     & 17.85$_{-0.73}^{+0.14}$ & 17.10$_{-0.70}^{+0.40}$ & 19.85$_{-0.15}^{+0.13}$ & 20.08$_{-0.09}^{+0.21}$ & 15.16$_{-0.26}^{+0.38}$ & 20.52$_{-0.26}^{+0.14}$ & 18.67$_{-0.14}^{+0.37}$ & 19.00$_{-0.14}^{+0.19}$ & 20.40$_{-0.16}^{+0.13}$ & 20.20$_{-0.08}^{+0.10}$ & 20.34$_{-0.171}^{+0.131}$      \\
\noalign{\medskip}
 0 & 1  &  9  & 170.49   & 18.32$_{-0.56}^{+0.39}$ & 17.90$_{-0.30}^{+0.20}$ & 19.56$_{-0.27}^{+0.17}$ & 19.96$_{-0.48}^{+0.18}$ & 16.00$_{-0.19}^{+0.28}$   & 20.40$_{-0.30}^{+0.25}$ & 19.11$_{-0.41}^{+0.07}$ & 19.04$_{-0.19}^{+0.21}$ & 20.25$_{-0.20}^{+0.18}$ & 20.05$_{-0.08}^{+0.19}$ & 20.32$_{-0.154}^{+0.181}$      \\
\noalign{\medskip}
 0 & 2  &  5  & 509.897  & 17.99$_{-0.63}^{+0.35}$ & 17.40$_{-1.00}^{+0.40}$ & 16.60$_{-0.37}^{+0.19}$ & 16.20$_{-0.38}^{+0.69}$ & 15.58$_{-0.21}^{+0.28}$   & 19.18$_{-0.26}^{+0.14}$ & 17.69$_{-0.39}^{+0.14}$ & 17.45$_{-0.21}^{+0.25}$ & 18.70$_{-0.22}^{+0.14}$ & 18.30$_{-0.30}^{+0.15}$ & 18.82$_{-0.241}^{+0.263}$       \\
\noalign{\medskip}
 0 & 3  &  21 & 1015.069 & 17.98$_{-0.61}^{+0.39}$ & 17.40$_{-0.30}^{+0.20}$ & 16.65$_{-0.20}^{+0.44}$ & 16.04$_{-0.55}^{+0.42}$ & 15.83$_{-0.11}^{+0.21}$   & 18.20$_{-0.24}^{+0.28}$ & 16.59$_{-0.31}^{+0.35}$ & 16.42$_{-0.35}^{+0.63}$ & 16.85$_{-0.25}^{+0.15}$ & 16.34$_{-0.18}^{+0.34}$ & 17.30$_{-0.830}^{+0.571}$     \\
\noalign{\medskip}
 0 & 4  &  9  & 1681.281 & 16.19$_{-0.59}^{+0.52}$ & 16.40$_{-0.80}^{+0.40}$ & 15.75$_{-0.30}^{+0.30}$ & 15.78$_{-0.84}^{+0.37}$ & 15.13$_{-0.07}^{+0.07}$   & 16.98$_{-0.25}^{+0.21}$ & 15.26$_{-0.47}^{+0.34}$ & 14.50$_{-0.37}^{+0.39}$ & 14.95$_{-0.16}^{+0.31}$ & 14.88$_{-0.27}^{+0.33}$ & 15.70$_{-0.186}^{+0.278}$     \\
\noalign{\medskip}
 0 & 5  & 33  & 2502.234 & 16.70$_{-0.42}^{+0.56}$ &            --         & 15.30$_{-0.40}^{+0.52}$ &            --           & 15.51$_{-0.06}^{+0.08}$     & 16.70$_{-0.35}^{+0.47}$ &     --                  & 14.50$_{-0.26}^{+0.15}$ & 14.75$_{-0.17}^{+0.13}$ & 14.50$_{-0.58}^{+0.34}$ & 15.60$_{-0.281}^{+0.494}$  \\
\noalign{\medskip}
 0 & 6  & 13  & 3470.052 &            --           &            --           & 13.45$_{-0.35}^{+0.38}$ &            --           & 14.69$_{-0.10}^{+0.11}$   & 15.63$_{-0.54}^{+0.41}$ &     --                  & 13.50$_{-0.19}^{+0.54}$ & 14.74$_{-0.24}^{+0.08}$ & 14.25$_{-0.51}^{+0.26}$ & 14.26$_{-0.260}^{+0.124}$        \\
\noalign{\medskip}
 0 & 7  & 45  & 4575.286 &            --           &            --           &          --            &            --           & 14.87$_{-0.08}^{+0.10}$    & 15.26$_{-0.56}^{+0.62}$ &     --                  & 13.18$_{-0.35}^{+0.42}$ & 14.15$_{-0.35}^{+0.15}$ & 14.27$_{-0.25}^{+0.31}$ & 13.99$_{-0.359}^{+0.236}$     \\
\noalign{\medskip}
 0 & 8  & 17  & 5806.903 &            --           &            --           &            --           &            --           & 13.89$_{-0.25}^{+0.18}$  & 13.28$_{-0.53}^{+0.66}$ &     --                  & 12.92$_{-0.38}^{+0.24}$ & 13.53$_{-1.42}^{+0.23}$ &    --                   & 12.60$_{-0.451}^{+0.721}$        \\
\noalign{\medskip}
 0 & 9  & 57  & 7152.314 &            --           &            --           &            --           &            --           & 14.31$_{-0.07}^{+0.08}$  & 13.53$_{-0.51}^{+0.24}$ &     --                  & 13.08$_{-0.26}^{+0.45}$ & 13.80$_{-0.52}^{+0.29}$ &    --                   &    --                      \\
\noalign{\medskip}
 0 & 10 & 21  & 8597.343 &            --           &            --           &            --           &            --           & 13.47$_{-0.55}^{+0.25}$   &      --                 &     --                  & 13.02$_{-0.33}^{+0.28}$ &    --                   &    --                   &    --           \\
\noalign{\medskip}
 1 & 0  & 1   & 5986.847 &            --           &            --           &           --            &            --           &           --          & 14.44$_{-0.58}^{+0.53}$ &     --                  & 14.28$_{-0.82}^{+0.30}$ & 13.45$_{-0.17}^{+0.12}$ &    --                   & 13.30$_{-0.380}^{+0.239}$     \\
\noalign{\medskip}
 1 & 1  & 9   & 6149.431 &            --           &            --           &           --            &            --           & 13.87$_{-0.06}^{+0.05}$  & 15.09$_{-0.59}^{+0.39}$ &     --                  & 14.34$_{-0.28}^{+0.28}$ & 14.07$_{-0.23}^{+0.03}$ &    --                   & 13.68$_{-0.223}^{+0.190}$    \\
\noalign{\medskip}
 1 & 2  & 5   & 6471.722 &            --           &            --           &           --            &            --           & 13.45$_{-0.11}^{+0.09}$   & 14.05$_{-0.28}^{+0.66}$ &     --                  &     --                  & 14.02$_{-0.19}^{+0.08}$ &    --                   & 13.57$_{-0.295}^{+0.191}$ \\
\noalign{\medskip}
 1 & 3  & 21  & 6950.843 &            --           &            --           &           --            &            --           & 13.91$_{-0.04}^{+0.04}$    & 14.74$_{-0.51}^{+0.27}$ &     --                  &     --                  & 14.23$_{-0.16}^{+0.10}$ &    --                   & 13.65$_{-0.386}^{+0.197}$   \\
\noalign{\medskip}
 1 & 4  & 9   & 7583.915 &            --           &            --           &           --            &            --           & 13.33$_{-0.09}^{+0.07}$    & 14.52$_{-0.43}^{+0.60}$ &     --                  &     --                  & 14.09$_{-0.21}^{+0.08}$ &    --                   & 13.00$_{-0.520}^{+0.253}$        \\
\noalign{\medskip}
 1 & 5  & 33  & 8363.744 &            --           &            --           &           --            &            --           & 13.89$_{-0.08}^{+0.08}$  & 13.70$_{-0.56}^{+0.34}$ &     --                  &     --                  & 13.96$_{-0.17}^{+0.01}$ &    --                   &    --           \\
\noalign{\medskip}
\hline
\end{tabular}
\label{results2}
\end{center}
\end{sidewaystable*}
}


Using the column densities measured for each level, we plotted the
excitation diagrams. These diagrams show the ratio of the column
density to the statistical weight of each energy level versus the
energy of the level. They characterize the excitation conditions of
the gas and give information about the temperature of the gas and its
physical conditions.  For the stars in our sample, the diagrams show
marked differences between the Herbig Ae/B9 stars, on the one hand,
and the Herbig Be stars, on the other. We distinguished between these
two sub-groups of stars in our analysis (see Sects.~\ref{disks} and
\ref{HBe}).
 
\subsection{ H$_2$ Modeling } 
\label{PDR}

Absorption line spectroscopy does not allow us to directly measure the
distance of the absorbing material along the line of sight.  In a
first attempt to better diagnose the origin of the detected molecular
gas and its excitation conditions, we modeled the CS H$_2$ spectra
using a photodissociation region (PDR) model. We used the
PhotoDissociation Region code of \cite{LePetit06}, referred to as the
''Meudon PDR Code". The code has been developed as a tool to
investigate the physical conditions in both diffuse and dense IS
clouds. The physics included in the code, as well as the numerical
techniques and several recent improvements, are fully described in
\cite{LePetit06}. We refer the reader to this paper and provide only
the main outlines of the code here.

The code models a slab of IS medium, in a stationary state, heated by
one star's radiation field and the galactic field in a plane-parallel
geometry. The IS radiation field flux is assumed isotropic on both
sides of the cloud, while the added incident stellar radiation is
normal on one side.  The code uses a standard galactic field law
\citep{Draine78}. The stellar FUV radiation field is given by a
blackbody law calculated from the star's radius and temperature and
the star-to-cloud distance. The model input parameters are thus the
radius and effective temperature of the star, the distance of the
observed gas from the star, the density of the medium, and the
extinction (\Av). The equations of thermal and chemical balance as a
function of depth within the cloud are solved in a plane-parallel
geometry with 113 chemical species and 967 chemical reactions
included. As a result, the code provides the temperature distribution
within the gaseous medium and the abundances of the chemical species
derived from the chemical balance.

Without any precise information on the nature of the dust grains in
the absorbing medium, we used standard ISM values for the grain size
distribution and properties \citep{Mathis77}. When available, we used
\iue\ spectra to determine the column density of \ion{H}{i}, which
allowed us to estimate the ratio of the gas column to the \Av\ as an
input parameter. We used the standard ISM value of this ratio when no
observational constraints were available.  For most of our targets,
little or no information is available in the literature about the
properties of the observed gas, namely its distance to the star or its
density and temperature. We thus proceeded by an iterative method,
varying both the distance and density of the gas to obtain the best
fits to the observed excitation diagrams. We checked that the model
output parameters are consistent with the molecular fraction we
derived from our spectral analysis (see Sect.~\ref{HBe}). We also
compared the synthetic spectra provided by the code to the observed
ones, as an additional check. The results of our modeling are
presented in the next sections.


\section{Ae/B9 stars}
\label{disks}

In this section, we first present the stars for which we have not
detected H$_2$ absorption lines. In the following subsections, when
H$_2$ is detected, we distinguished the stars for which we observed
cold H$_2$ and warm H$_2$ on the basis of their excitation diagrams.

\subsection{Stars lacking H$_2$ absorption lines}

We could not detect any H$_2$ absorption lines for 6 stars in our
sample. \\

\begin{figure}[!htbp] 
   \centering
   \includegraphics[width=9cm]{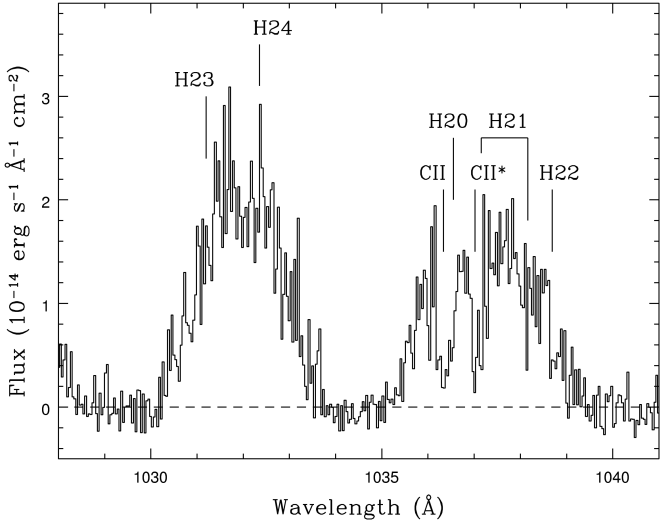}
   \caption{\ion{O}{vi} emission doublet near 1032 \AA\ and 1038 \AA\
     in the \fuse\ spectrum of $\beta$ Pictoris observed in
     2002. Here, only the exposures obtained during the night have
     been co-added, to avoid contamination by strong airglow emission
     lines due to the Earth's atmosphere. The positions of the H$_2$
     transitions we should observe if H$_2$ was present along the line
     of sight are plotted. The H$_2$ levels are marked H20 for
     ($v=0,J=0$), H21 for ($v=0,J=1$), etc. For details on the
     \ion{C}{ii} absorption, see \cite{Roberge06}.}
         \label{H2_BPic}
\end{figure}

{\bf \bpic}: Early \fuse\ observations of \bpic\ in 2000 allowed us to
report a clear deficiency of molecular hydrogen in the disk of the
star, consistent with the evolved status of the system
\citep{LECAV01}. Taking advantage of a third exposure acquired in
November 2002 with a total exposure time of about 50\,000 seconds,
much longer than the two previous ones, we checked for any H$_2$
absorption features in the new spectra (Fig.~\ref{H2_BPic}).
Co-adding all three spectra allowed us to refine the upper limit on
the total H$_2$ column density previously estimated at about 10$^{18}$
cm$^{-2}$ by \cite{LECAV01}. Assuming a low value for the intrinsic
line width ($b=1$~\kms), we found the new estimate to be less than
2.6$\times$10$^{17}$ cm$^{-2}$.

{\bf HD~109573}: HD~109573 is the only star in our study for which no
H$_2$ absorption lines could be detected, despite a noticeable stellar
continuum flux. The stellar flux is well-developed in the reddest part
of the \fuse\ wavelength range, above 1110~\AA. Below this wavelength,
where the strongest H$_2$ transitions fall, the stellar flux level is
much lower but would be high enough to observe H$_2$ absorption lines
if present, but the signal-to-noise ratio is very low (S/N $\simeq$
1.5) and no stellar emission lines are present to overcome this issue.

HD~109573 is also an evolved system, the star being very close to the
main sequence. It is known to possess a flat debris disk seen nearly
edge-on, with an inclination angle of about 20{\degr} from the line of
sight \citep{Augereau99}. The dusty disk has two components: an inner
annulus that extends between 4--9 AU to $\sim$55 AU from the central
star, and an outer ring at 70 AU \citep{Wahhaj05, Augereau99}. These
two components are separated by a gap located between 55 and 60 AU
from the star, which is probably the consequence of planet formation
\citep{Augereau99}. The outer ring is about 15 AU thick, with an
inclination angle of 13$\pm$1{\degr} from the line of sight
\citep{Telesco00}. This implies that the upper layers of the disk
intersect the line of sight and could potentially produce absorption
features.  However, as previously reported by \citet{Chen04}, we do
not observe absorption lines of H$_2$ in the \fuse\ spectrum.  We set
an upper limit on the total column density of H$_2$ of $\leq 2.5\times
10^{15}$ cm$^{-2}$, which is consistent with the limit found by
\citet{Chen04}. This non-detection of H$_2$ implies that the disk of
HD~109573 is almost depleted of molecular gas, which agrees with the
evolutionary status of this star. \\

{\bf NX~Pup, HD~36112, HD~135344, and HD~100453}: The photospheric
continuum of these four stars is too low in the FUV (S/N $\sim$2 at
1100~\AA) for a reliable detection of CS/IS absorption lines. However,
as in the case of \bpic, we took advantage of the \ion{O}{vi} emission
lines to place limits on the total H$_2$ column densities along the
line of sight. The presence of a CS disk around NX~Pup is suspected
from its SED. For the three other stars, all younger than \bpic, the
non-detection of H$_2$ is consistent with the high inclination angles
for their disks, which do not intersect the lines of sight to the
stars (see Table~\ref{tab_sample}). It shows that, if present, the CS
gas does not extend far enough above the midplanes of the disks to be
detected in absorption and that the remnant of the initial CS
envelopes have been dispersed or are too low in density to allow
detection.

\subsection{Cold H$_2$}
\label{Cold_H2}

{\bf HD~141569}: The HD~141569 disk is usually called a transitional
disk, in which there are signs of dust clearing in the inner disk, but
significant primordial molecular gas is still supposed to be present
in the disk. In a previous detailed analysis \citep[see][]{klr05}, we
demonstrated that the absorption lines of molecular hydrogen and
several atomic species observed in its \fuse\ spectrum are very likely
related to the slightly reddened diffuse outer region surrounding the
L134N dark cloud complex \citep{Juvela02}.  The non-detection of CS
gas implies that there is no remnant CS envelope and that all the gas
has had time to collapse into a flat or very slightly flared
disk. This result is consistent with the $^{12}$CO observations at
345.796 GHz described in \citet{Dent05}. \\

{\bf AB Aurig\ae}: The presence of H$_2$ in the far-UV spectra of AB
Aurig\ae\ was first reported and analyzed by \citet{ROBERGE01} from an
early \fuse\ observation. A new observation obtained more than one
year later allows us to confirm and refine the analysis.

\begin{figure}[!htbp] 
   \centering
   \includegraphics[width=9cm]{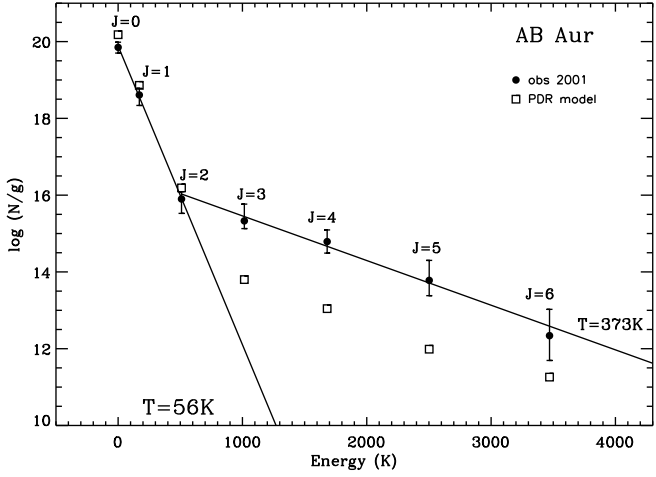}
   \caption{Excitation diagram for H$_{2}$ towards AB Aurig\ae. The
     observed level column densities (filled circles) show that H$_2$
     is thermalized up to $J=2$ with a low kinetic temperature of
     about 56~K, while the column densities of the higher J-levels are
     consistent with a temperature of about 373~K. The populations of
     the first three energy levels are reproduced well by a PDR model
     (open square) with a low density of about 250 cm$^{-3}$ and a
     distance between the star and the gas of about 0.3 pc, with an
     excitation temperature of about 55~K. Our model does not
     reproduce the higher energy levels' excitation. This may be due
     to the presence of a second warm/hot gaseous component along our
     line of sight (see text).}
         \label{AB_Aur}
\end{figure}

We identified and measured H$_2$ lines arising from $J=0$ to $J=6$ in
the ground vibrational state ($v=0$). As suggested by
\cite{ROBERGE01}, the H$_2$ could be IS in origin or have a CS origin,
as a cold remnant of its natal cloud or a CS envelope
\citep{Semenov05}. The relatively low spectral resolution of \fuse\
prevents us from drawing firm conclusions about the origin of the
detected gas on the basis of its radial velocity and line width. The
excitation diagram derived from the column densities in the J-levels
may provide additional clues for further constraining the location of
the observed gas. For AB~Aur, it confirms that the H$_2 $ is
thermalized up to $J=2$ with a low kinetic temperature of 56$\pm$4~K
(Fig.~\ref{AB_Aur}), results in good agreement with those of
\citet{ROBERGE01}.  Since the relative populations of the lower H$_2$
energy levels are determined primarily by thermal collisions, the
excitation temperature found from these levels should be close to the
kinetic temperature of the gas. On the other hand, the column
densities of the higher J-levels are compatible with thermal
equilibrium at a temperature of 373$\pm$83~K.

Using the Meudon PDR Code, we checked that such excitation conditions
could not be reproduced by a single absorbing component along the line
of sight towards the star. We varied the different parameters within
their error bars and tried to model the observations with different
\Av\ values found in the literature, but no PDR model fit the
data. There are still some free parameters to explore, but we need
additional observational constaints to make an accurate modeling. As
shown in Fig.~\ref{AB_Aur}, a diffuse component
(n$_H$$\sim$250~cm$^{-3}$) located at 0.3 pc from the star reproduces
the column densities of the three first energy levels, but failed to
reproduce the higher J-level populations.

Two-temperature behavior in excitation diagrams for H$_2$ is often
observed in diffuse and translucent IS clouds
\citep[e.g.][]{Spitzer73, Gry02, Browning03}, but its origin is not
understood well and remains a subject of controversy, as reviewed and
discussed in \cite{Snow06}. Several excitation mechanisms could be
invoked separately or all together to explain this behavior.
\cite{Shull82} initially suggested that the populations of the higher
levels can be inflated by UV and/or formation pumping and radiative
cascades. We checked that, to populate the $J>2$ levels by UV pumping
alone, we would have to increase the radiation field (G$\sim$1.5 in
Draine's units) by a factor of more than 10. An alternative model
invokes local heating of the gas within the cloud. Such heating could
be generated by magneto-hydrodynamic shocks \citep[see][]{Chieze98} or
transient small-scale turbulence that results in localized tiny warm
regions, as suggested by \cite{Falgarone95}.

The simplest explanation is, however, the presence of distinct
absorbing media along the line of sight with very different excitation
conditions \citep{Browning03}. For all the stars in this study, we
found $b$-values much higher than expected for pure thermal
broadening. This is an unusual situation in diffuse ISM clouds, where
typical $b$-values are about 1.5 up to 3 \kms\
\citep[e.g.][]{Gry02}. However, higher values are also routinely
reported \citep[e.g.][]{Lehner03, Redfield04}. The simplest
explanation for this phenomenon is the presence of several gaseous
components along the line of sight that could appear blended in the
spectra. But it could also be related to gas motions on different
spatial scales as suggested by different studies. Such motions could
originate in turbulent and warm layers within the cloud, through
transient shocks or vortices \citep{Joulain98, Gredel02}, which lead
to increased observed $b$-values. Without observations at higher
spectral resolution or additional constraints on the chemistry of the
gaseous medium, we cannot distinguish between explanations.

In conclusion, we cannot distinguish with certainty yet whether the
H$_2$ gas in this source has a circumstellar or interstellar origin,
but the cold component appears to be located at a relatively large
distance from the star ($\sim$0.3 pc).

Now the question arises, whether the gas has a circumstellar origin,
i.e. could it be part of the disk or must it be part of a separate,
more spherically distributed, components of CS gas? AB Aur is known to
harbor an extended CS disk with an inclination angle between 27{\degr}
and 35{\degr} from the plane of the sky \citep{Eisner03, Pantin05},
i.e.\ nearly face-on. Since a detection of H$_2$ in absorption means
that the line of sight toward the star must pass {\em through} the
H$_2$ gas, the near-face-on inclination appears to make it unlikely
that the gas is part of the disk. However, a more quantitative
analysis is required to make any firm statements.  Suppose the disk is
at its thickest (geometrically) at a radius $r$, where it has a gas
surface density $\Sigma_{\mathrm{gas}}$ and a gas temperature
$T$. Assuming vertical hydrostatic equilibrium, the gas density as a
function of height $z$ above the midplane at that radius is

\begin{equation}\label{eq-vert-struct}
\rho_{\mathrm{gas}}=\frac{\Sigma_{\mathrm{gas}}}{\sqrt{2\pi}H_p}
\exp\left(-\frac{z^2}{2H_p^2}\right)
\end{equation}

with

\begin{equation}
H_p = \sqrt{\frac{kTr^3}{\mu m_p G M_{*}}}~~~,
\end{equation}

{\noindent}where $\mu=2.3$ is the mean molecular weight of the
molecular-hydrogen-helium mixture, $m_p$ the proton mass, $G$ the
gravitational constant, and $k$ the Boltzmann constant. The column
depth along the line of sight passing through $(r,z)$, i.e.\ with the
disk at an inclination of $i = \arctan (r/z)$ away from face-on, is
then estimated as

\begin{equation}\label{eq-vert-struct-2}
N_{\mathrm{H}_2} = \frac{\rho_{\mathrm{gas}}l}{\mu m_p}
\simeq \frac{\rho_{\mathrm{gas}}H_p}{\mu m_p}~~~.
\end{equation}

{\noindent}Here we simply took as radial extent $l$ the pressure scale
height $l\simeq H_p$. In reality this may be a factor of a few higher.

Let us take $r$ to be the outer radius of the disk estimated at
$r_{\mathrm{out}}=$800 AU \citep{Semenov05} and a disk mass of at most
1 $M_{\odot}$.  With $M_{\mathrm{disk}}\simeq
\Sigma_{\mathrm{gas}}(r_{\mathrm{out}})\pi r_{\mathrm{out}}^2$, this
gives a surface density of about $\Sigma_{\mathrm{gas}}\simeq 4$
g/cm$^2$ at $r=r_{\mathrm{out}}$. Using these numbers and the stellar
parameters from Table~\ref{tab_sample}, we find
$H_p(r_{\mathrm{out}})/r_{\mathrm{out}}=0.36$ for a temperature of
$T=95$~K, which, with an inclination of 35$\degr$ away from face-on,
yields a column density of $\log N(\mathrm{H}_2)=20$ as given in
Table~\ref{results}. In other words: we need a temperature of about
95~K to puff the disk up enough for sufficient matter to cross the
line of sight at around 800 AU distance. Considering the roughness of
this estimate and the closeness to the excitation temperature of 56~K
for the low-lying levels, it shows that it is possible in principle
that this gas is part of the outer part of a massive disk. But the
same derivation yields a volume density of $n_{\mathrm{H}_2}\simeq
3.4\times 10^4$, which is much higher than the $\sim 250$ derived from
the PDR model. Instead of 800 AU, we could also use $0.3$ pc $\equiv
61900$ AU as reference radius $r$. Then any estimate would yield the
disk as gravitationally unbound, and we cannot speak of a
``disk''. Thus, if the cool material is indeed located at such a
distance, it may still be associated with the star AB Aurigae, but it
would be meaningless to speak of disk material there and is likely to
be either a remnant envelope or outflowing material from the disk.

\subsection{Warm H$_2$ gas}
\label{Ae}

\begin{figure}[!htbp] 
   \centering \includegraphics[width=9cm]{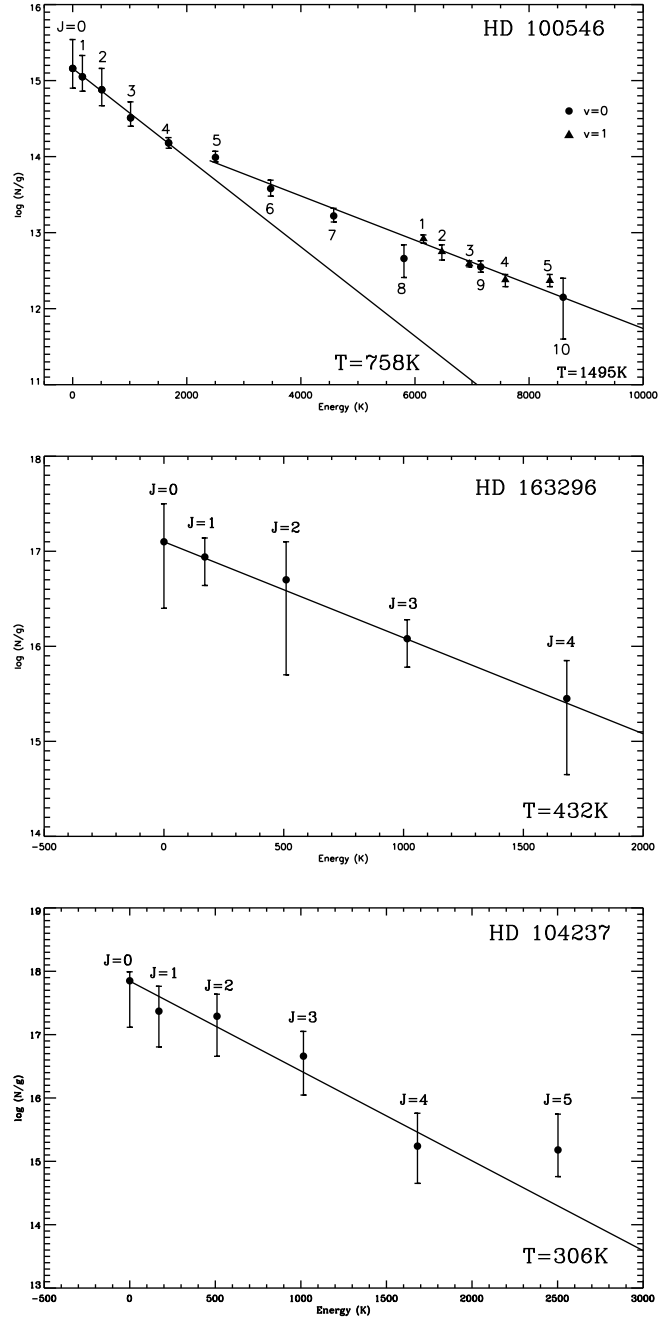}
   \caption{(a) Excitation diagram of H$_2$ for HD~100546. The H$_2$
     is thermalized up to $J=4$ with a kinetic temperature of
     758$\pm$147~K. The column densities of the higher J-levels are
     consistent with a temperature of about 1495$\pm$70~K. (b)
     Excitation diagram of H$_2$ for HD~163296. As for HD~100546, the
     H$_2$ is thermalized up to $J=4$ with a lower kinetic temperature
     of 432$\pm$135~K. Here, only the $v=0$ levels up to $J=4$ are
     observed because of the later type of the star (A1Ve) and the low
     S/N ratio of the spectrum. (c) Excitation diagram of H$_2$ for
     HD~104237. The H$_2$ is thermalized up to $J=4$ with a kinetic
     temperature of 306$\pm$80~K. The higher J-levels may be
     consistent with a higher temperature, as suggested by the $J=5$
     column density.}
         \label{HD100_HD163}
\end{figure}

We re-did the spectral analyses for HD~100546 and HD~163296, which
were presented in \citet{LECAV03}, and found the same results. We also
refined the analysis of the \fuse\ spectrum of HD~104237 presented in
\cite{Herczeg03} and found compatible results. In the HD~100546
spectrum, lines arising from the J-levels of the ground vibrational
level ($v=0$) up to $J=10$ and J-levels of the first vibrational level
($v=1$) up to $J=5$ are detected. In the HD~163296 and HD~104237
spectra, only lines arising from the J-levels up to $v=0$, $J=4$ and
$v=0$, $J=5$, respectively, are observed. Higher rotational J-levels
and vibrational levels are beyond the detection limit, because of the
sharp flux decrease with decreasing wavelength caused by the later
spectral types of these two stars and because of the lower
signal-to-noise ratios of the data. These low signal-to-noise ratios
are responsible for the rather large uncertainties in the
determinations of the column densities (Fig.~\ref{HD100_HD163}) .

We plotted the excitation diagrams of H$_2$ for each of the three
stars, and found two-temperature behavior in the case of HD~100546
(Fig.~\ref{HD100_HD163}). In contrast to the sources in
Sect.~\ref{Cold_H2}, for all three stars, the H$_2$ is thermalized up
to $J=4$ with surprisingly high kinetic temperatures, greater than
300~K and as high as 758~K in the case of HD~100546
(Fig.~\ref{HD100_HD163}). In addition, the temperature derived for the
higher J-levels observed in HD~100546, is 1495$\pm$70~K. Such
excitation conditions clearly differ from those observed in the
interstellar medium. In addition, the measured radial velocities
favor a CS origin for the H$_2$ at the \fuse\ spectral resolution.

Using the numerical model developed by \cite{LEBOURLOT93}, we
estimated that the warm H$_2$ (with $J \leq 4$) should be located at a
distance of about 1.5~AU from HD~100546, 4~AU from HD~163296, and
9.5~AU from HD~104237 to explain the observed excitation conditions
\citep[for details about the method see ][]{LECAV03}. The observed
temperature cannot be explained by the Galactic UV radiation field,
which implies that the observed H$_2$ is circumstellar and lies very
close to its central star. Using the volume densities derived from the
critical density given by \cite{LEBOURLOT99}, we found typical
thicknesses of $\leq$0.2~AU for the absorbing layer towards HD~100546,
$\leq$1.2~AU for HD~163296 and 3.2~AU for HD~104237. These results
reinforce our interpretation of a CS origin for the detected gas.

This again poses the question of whether this material is part of the
disk or part of some more spherically distributed gaseous
circumstellar material. For all three sources, the inclination is more
than 50{\degr} away from edge-on. Using Eq.~\ref{eq-vert-struct-2}
with a gas surface density of, say, $\Sigma_{\mathrm{gas}}=10^4$
g/cm$^2$, at the radii estimated above, and inclination of 50{\degr}
and required column densities given in Table~\ref{tab_sample}, we find
that we need a temperature of $11600$~K for HD~100546, $4700$~K for
HD~163296, and $2300$~K for HD~104237 to support hydrostatic
equilibrium and yet have the required column density of H$_2$. All
these temperatures are well above the excitation temperatures measured
for these sources, so the disk is excluded as the carrier of the warm
H$_2$ gas we detected.

We can think of two possibilities to explain the excitation of this
gas. The first is that this medium is related to a non-radiatively
heated medium close to the stellar surface, such as a chromosphere, as
proved by the well-known FUV and X-ray activity of these stars
\citep[e.g. ][]{D04, D05, Grady04}. Others Herbig Ae stars exhibit
noticeable activity in the far-UV but at a much lower level
\citep{D06}. The H$_2$ we observe could be produced in the very outer
layer of this extended hot region. Due to the inclination angle of the
disk, this region should have a more or less spherical geometry to
allow gas to be observed in absorption.

Another natural origin for the excited H$_2$ is photoevaporation of
the disk by FUV stellar radiation (Hollenbach private
communication). This is a mechanism of disk dissipation similar to the
better known EUV-driven photoevaporation \citep[e.g.][]{Hollenbach94},
but this time driven by non-ionizing FUV photons. In contrast to the
ionized EUV-driven wind, which originates as close as $\sim$1~AU from
the star, the FUV-driven wind is neutral and originates in the outer
regions of the disk (many tens of AU away from the star). It is
expected to have a much stronger wind mass loss rate than the EUV
wind. According to Gorti \& Hollenbach (ApJ submitted), such strong
disk winds are expected to flow away from the source in a nearly
spherical way, and they are therefore interesting candidates for
explaining the observed warm circumstellar matter. The problem in this
case is, however, that this FUV driven wind originates much farther
out than the few AU radii derived from the excitation diagrams.

Further observations, such as IR spectra of H$_2$, are required to
give a better understanding of the origin of the gas and the physical
processes that produce this excited H$_2$. More modeling of
photoevaporating gas or chromosphere models, which would take non-LTE
excitation mechanisms into account, could also help to better
constrain the excitation conditions.


\section{Be stars}
\label{HBe}
This second group of stars contains all the stars earlier than B9
type, which are also the youngest stars in the sample. For all but
one, namely HD~250550, we have identified and measured lines arising
from the rotational levels of the ground vibrational level, up to at
least $J = 7$, and also lines from the first vibrational level. The
derived excitation diagrams, plotted in Fig.~\ref{Be}, show that the
H$_2$ is thermalized up to $J=3$ with a kinetic temperature around
100~K. The temperatures given by the column densities of the high
J-levels range from $\sim$500~K to $\sim$1600~K. The radial velocities
of this matter suggests that it is of circumstellar origin.

As seen in Fig.~\ref{Be}, the excitation diagrams point to similar
excitation conditions for all the stars, which is a clue to a common
structure in their CS environments. Moreover, since the \vsini\ of
these stars span a wide range of values (see Table~\ref{tab_sample})
and since the detection of H$_2$ requires the line-of-sight to go
through the material, it is unlikely that the material in all these
objects is part of a circumstellar disk. Instead it is more likely
that this matter is part of the remnants of the clouds within which
the stars were formed.

\subsection{Modeling the H$_2$ excitation}

Little is known about the CS environments of these young stars and to
model the physical conditions of the gas we detected, we had to
estimate realistic values for the input parameters. To derive the
\ion{H}{i} column density for each star, we used archival \iue\
spectra to model the \lyalpha\ line (Table~\ref{gas-dust}), when good
data were available. Using the E(B-V) values listed in
Table~\ref{tab_sample}, we calculated the gas-to-dust ratios
N(H$_{tot}$)/E(B-V), in the environments of the stars and the
molecular fractions ($f$), i.e. the fraction of hydrogen atoms in
molecular form (Table~\ref{gas-dust}). Our gas-to-dust ratios values
are comparable to the mean interstellar value of about $5.8 \times
10^{21}$ atoms cm$^{-2}$ mag$^{-1}$, as measured for standard IS
clouds within 2 kpc of the Sun \citep{BOH78}. This favors an IS origin
for the H$_2$ we observed towards these stars.

\begin{table}[!h]
\begin{center}
  \caption{Column densities of \ion{H}{i} measured from the \iue\
    spectra, when good data were available.}
\begin{tabular}{llcccccc}

  \hline
  \hline
  Stars     & Column density                           &  N(H$_{tot}$)/E(B-V)                  & $f$ \\ 
            & of \ion{H}{i} (cm$^{-2}$)                &  (atoms c$m^{-2}$ mag$^{-1}$)          &  \\
  \hline                                        
  HD~250550 & 9.00$_{-4.0}^{+5.0}$$\times$10$^{20}$ (1)  & 4.26$_{-1.92}^{+2.33}$$\times$10$^{21}$ & 0.04$_{-0.03}^{+0.06}$ \\  
  HD~259431 & 1.60$_{-0.6}^{+2.4}$$\times$10$^{21}$ (1)  & 9.51$_{-3.49}^{+10.2}$$\times$10$^{21}$ & 0.35$_{-0.24}^{+0.37}$  \\    
  HD~38087  & 1.91$_{-0.41}^{+2.09}$$\times$10$^{21}$    & 8.46$_{-1.41}^{+7.9}$$\times$10$^{21}$  & 0.22$_{-0.13}^{+0.15}$ \\  
  HD~76534  & 2.40$_{-1.4}^{+1.6}$$\times$10$^{21}$      & 1.25$_{-0.63}^{+0.77}$$\times$10$^{22}$ & 0.27$_{-0.16}^{+0.51}$    \\
  \hline
\end{tabular}
\begin{list}{}{}
 (1) \cite{JCB03}
\end{list}
\label{gas-dust}
\end{center}
\end{table}

For each star, we run grids of models using the Meudon PDR Code, with
different distances and densities for the gas, to find the
best-fitting model that is consistent with the observed values of the
excitation temperature and molecular fraction. The best-fitting model
allowed us to derive the total density and the distance of the gas
from the central star, which are not constrained by absorption lines
observations. Our results are tabulated in Table~\ref{param_pdr}.

\begin{table}[!h]
\begin{center}
  \caption{Distance of the gas from the central star and total density
    of the observed medium, as determined by the best-fitting model
    using the Meudon PDR Code.}
\begin{tabular}{lccccccc}
\hline
\hline
Star        & Gas to star   & Gas  volume       \\ 
            & distance (pc) & density, n$_H$ (cm$^{-3}$)     \\
\hline                                       
HD~176386   & 0.04         & 3000     \\ 
HD~250550   & 0.03         & 600      \\  
HD~85567    & 0.10         & 250       \\  
HD~259431   & 0.35         & 70      \\    
HD~38087    & 1.60         & 12      \\         
HD~76534    & 0.40         & 170     \\     
\hline
\end{tabular}
\label{param_pdr}
\end{center}
\end{table}

  \begin{figure*}[!htbp] 
\vspace{1cm}
   \centering \includegraphics[width=18cm]{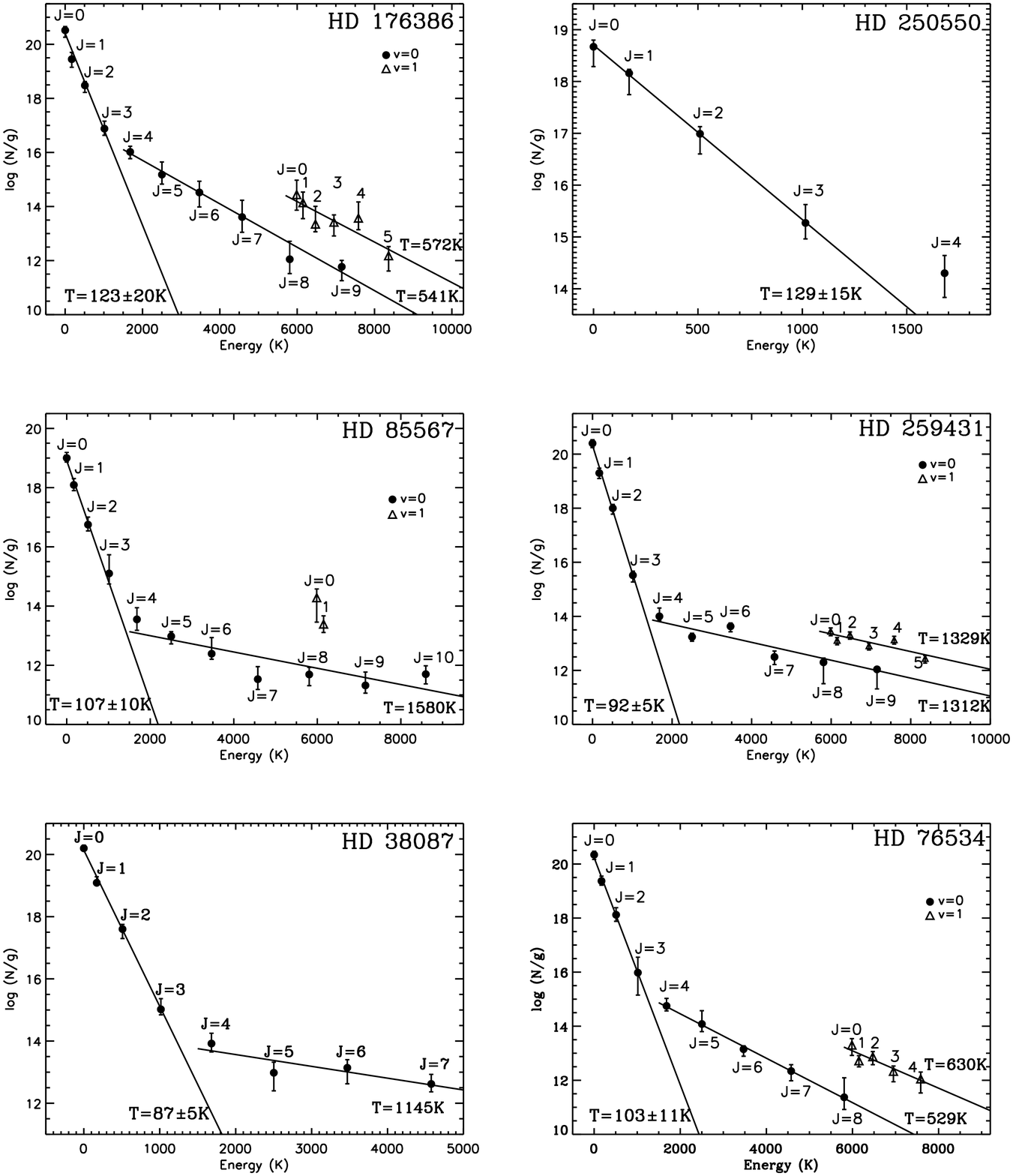}
   \caption{Excitation diagram for H$_2$ towards the Be stars in our
     sample. For each star, the H$_2$ is thermalized up to $J \geq 3$
     with a kinetic temperature of about $\sim$~100~K. \vspace{1cm}}
         \label{Be}
   \end{figure*}

  \begin{figure*}[!htbp] 
\vspace{1cm}
   \centering \includegraphics[width=18cm]{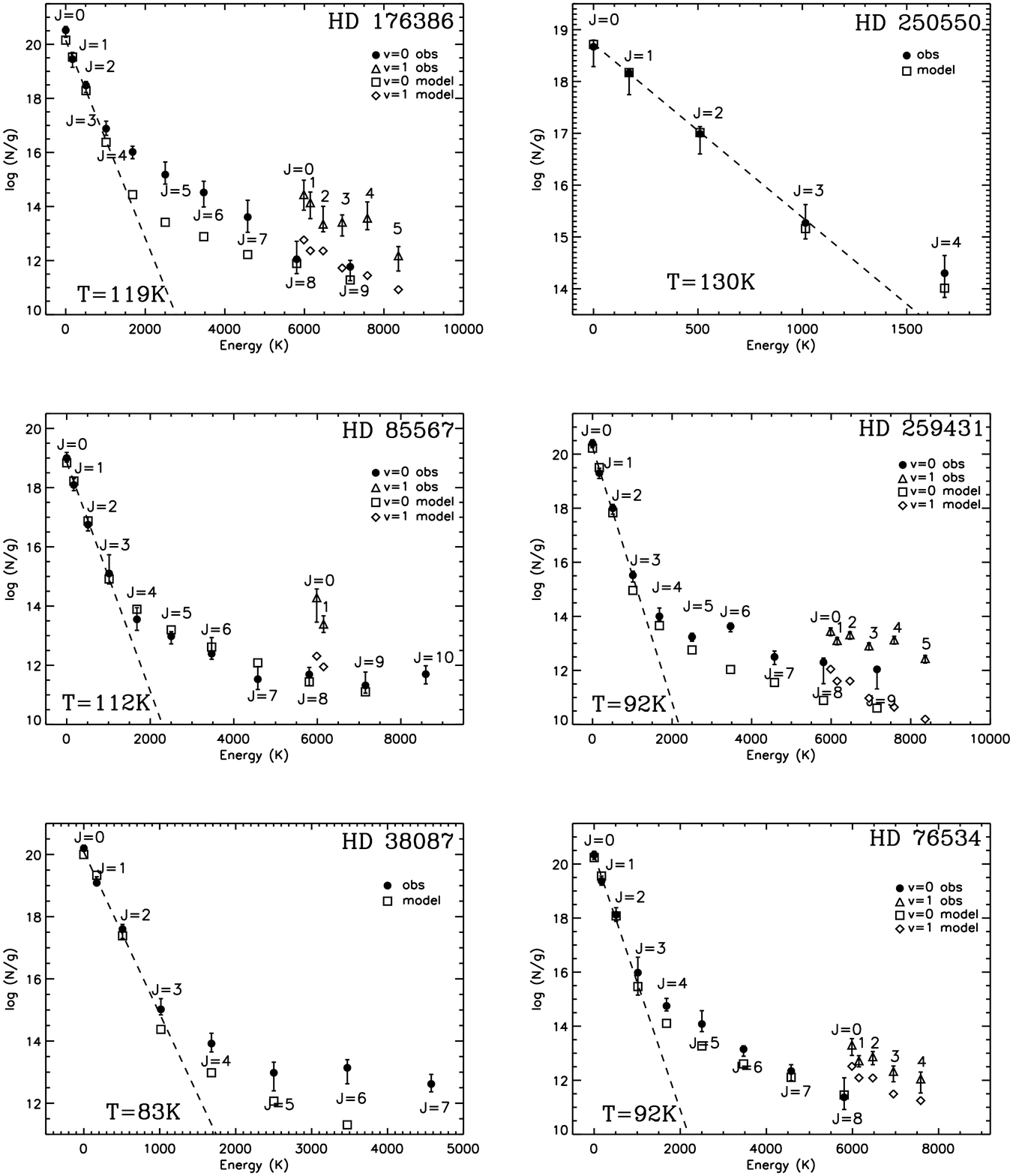}
   \caption{Best-fitting models obtained with the Meudon PDR Code
     overplotted on the excitation diagrams derived from the observed
     column densities (Fig.~\ref{Be}). The general shape of the
     diagrams are well-reproduced, but the excitation conditions of
     the higher J-levels cannot be reproduced well by the model (see
     text).  \vspace{1cm}}
         \label{PDR}
\end{figure*}


\begin{figure*}[!htbp] 
 \centering 
\vspace{2cm}
\includegraphics[width=16cm]{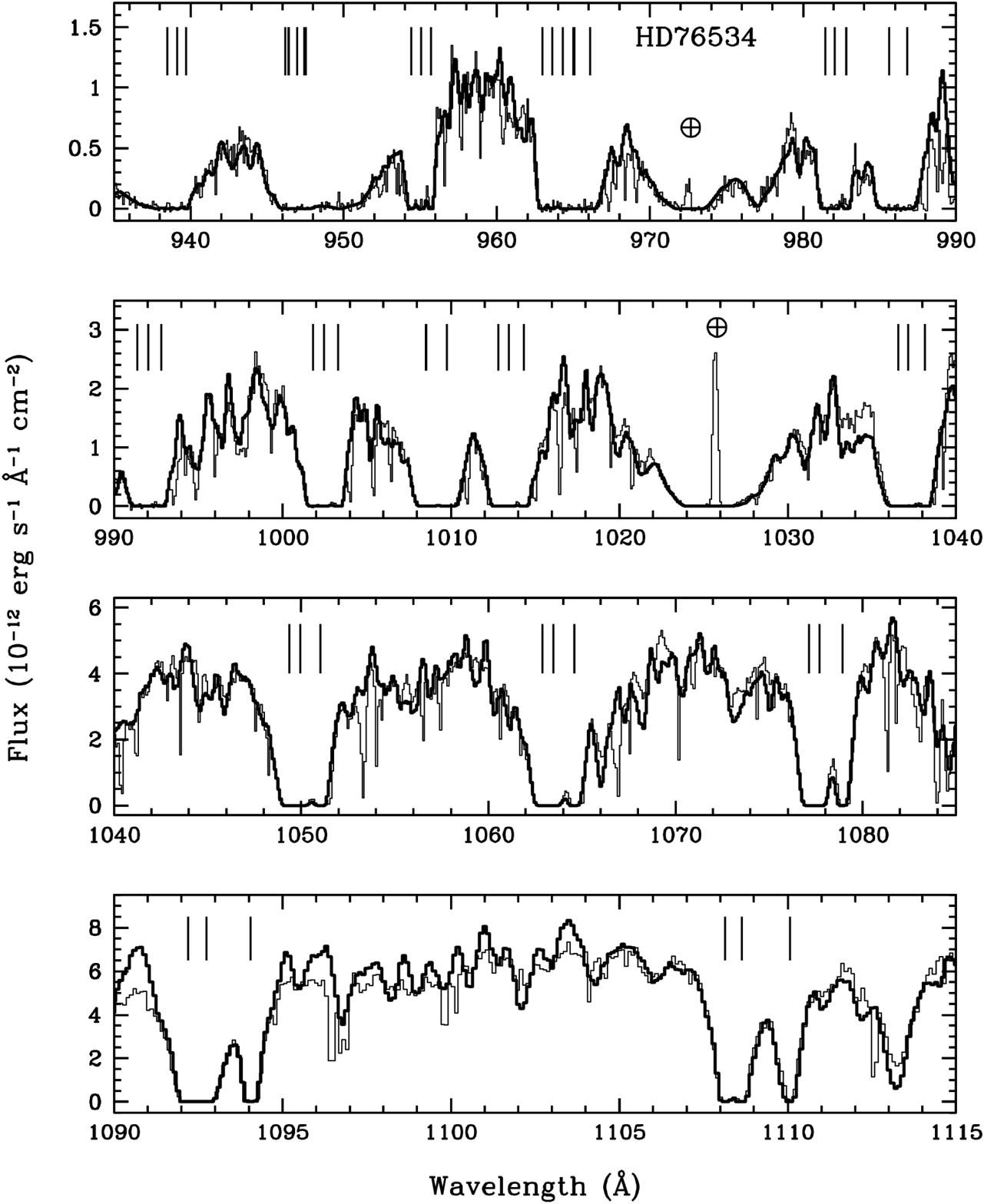}
\caption{ Overplot of the \fuse\ spectrum of HD~76534 (thin line) and
  the synthetic spectrum obtained by combining our photospheric model
  with the results of the Meudon PDR Code (thick line). This plot only
  presents the portions of the spectrum where the stronger absorption
  lines of H$_2$ ($v=0, J=0, 1, 2$) are observed. These principal
  prominent absorption troughs caused by molecular hydrogen are
  indicated throughout the spectrum. Airglow lines are indicated by
  $\oplus$ symbols. \vspace{2cm}}
      \label{spec_synt_hd76}
\end{figure*}

\onlfig{8}{
\begin{figure*}[!htbp] 
 \centering 
\vspace{2cm}
\includegraphics[width=16cm]{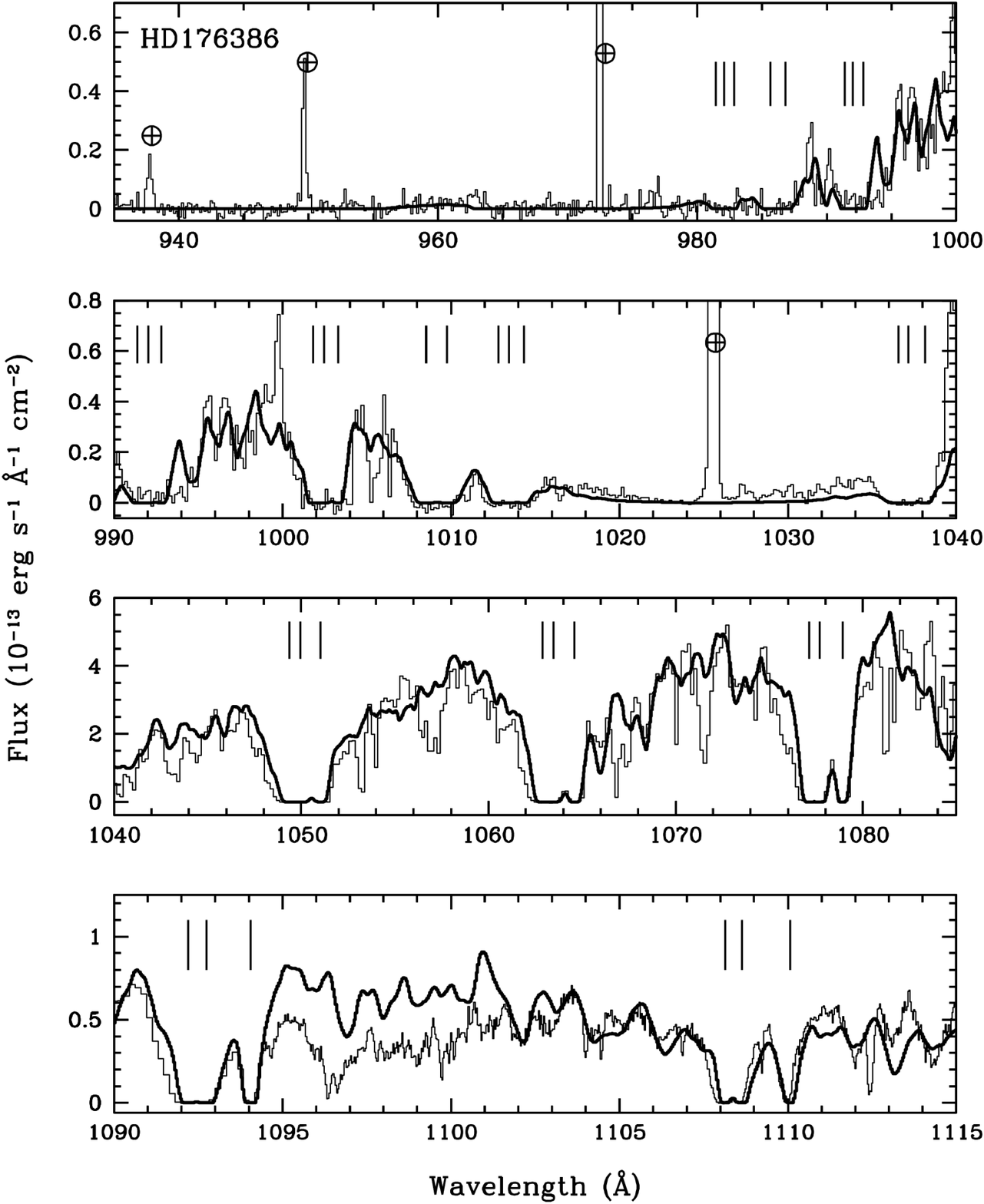}
\caption{Overplot of the \fuse\ spectrum of HD~176386 (thin line) and
  the synthetic spectrum obtained by combining our photospheric model
  with the results of the Meudon PDR Code (thick line). This plot only
  presents the portions of the spectrum where the stronger absorption
  lines of H$_2$ ($v=0, J=0, 1, 2$) are observed. These principal
  prominent absorption troughs caused by molecular hydrogen are
  indicated throughout the spectrum. Airglow lines are indicated by
  $\oplus$ symbols.\vspace{2cm}}
      \label{spec_synt_hd176}
\end{figure*}
}

\onlfig{9}{
\begin{figure*}[!htbp] 
 \centering 
\vspace{2cm}
\includegraphics[width=16cm]{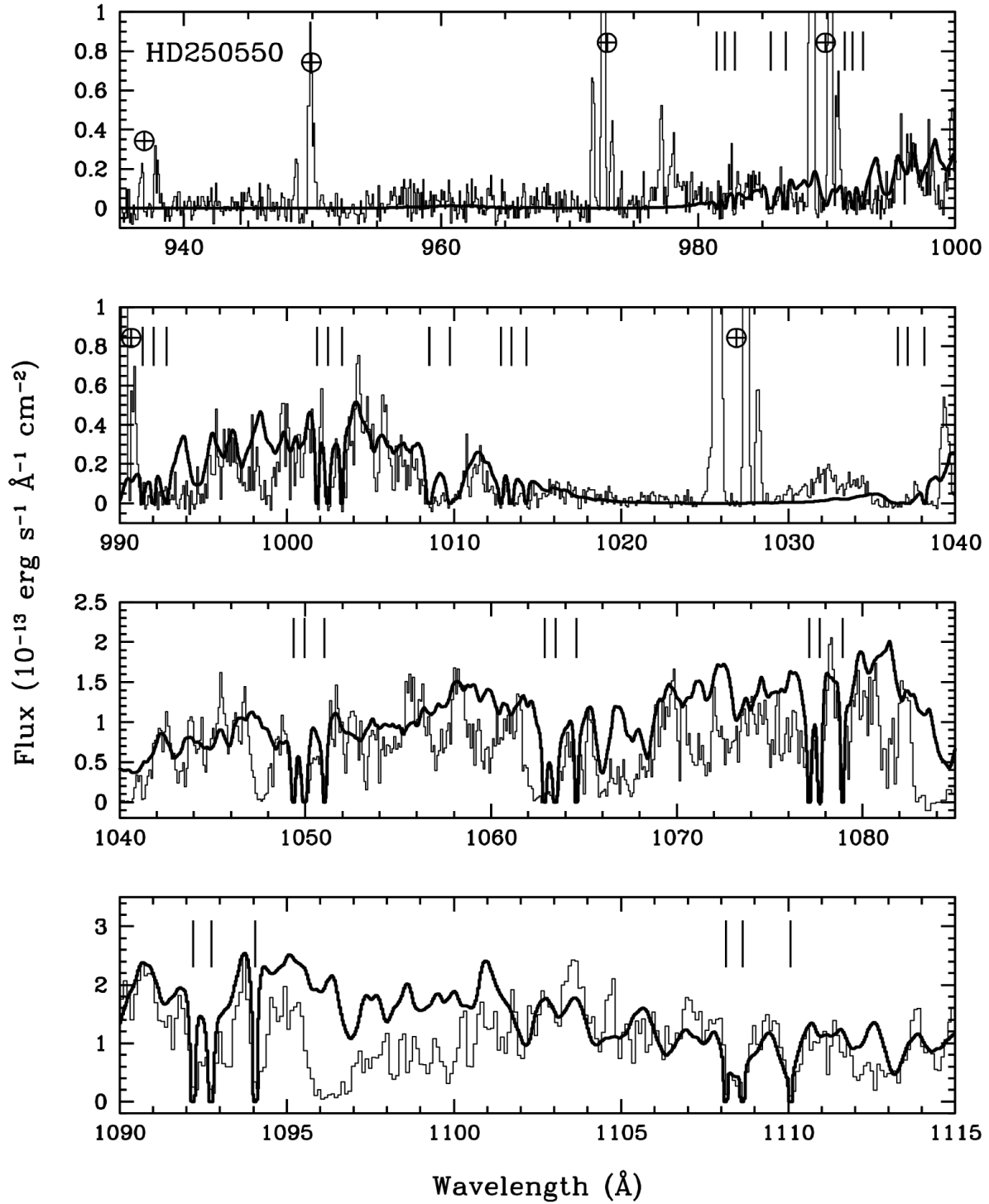}
\caption{Overplot of the \fuse\ spectrum of HD~250550 (thin line) and
  the synthetic spectrum obtained by combining our photospheric model
  with the results of the Meudon PDR Code (thick line). Same legend as
  for Fig.~\ref{spec_synt_hd176}. Here, the photospheric model does
  not fit the data well because of the strong stellar wind lines
  \citep[for details see ][]{JCB03}.\vspace{2cm}}
      \label{spec_synt_hd250}
\end{figure*}
}

\onlfig{10}{
\begin{figure*}[!htbp] 
 \centering 
\vspace{2cm}
\includegraphics[width=16cm]{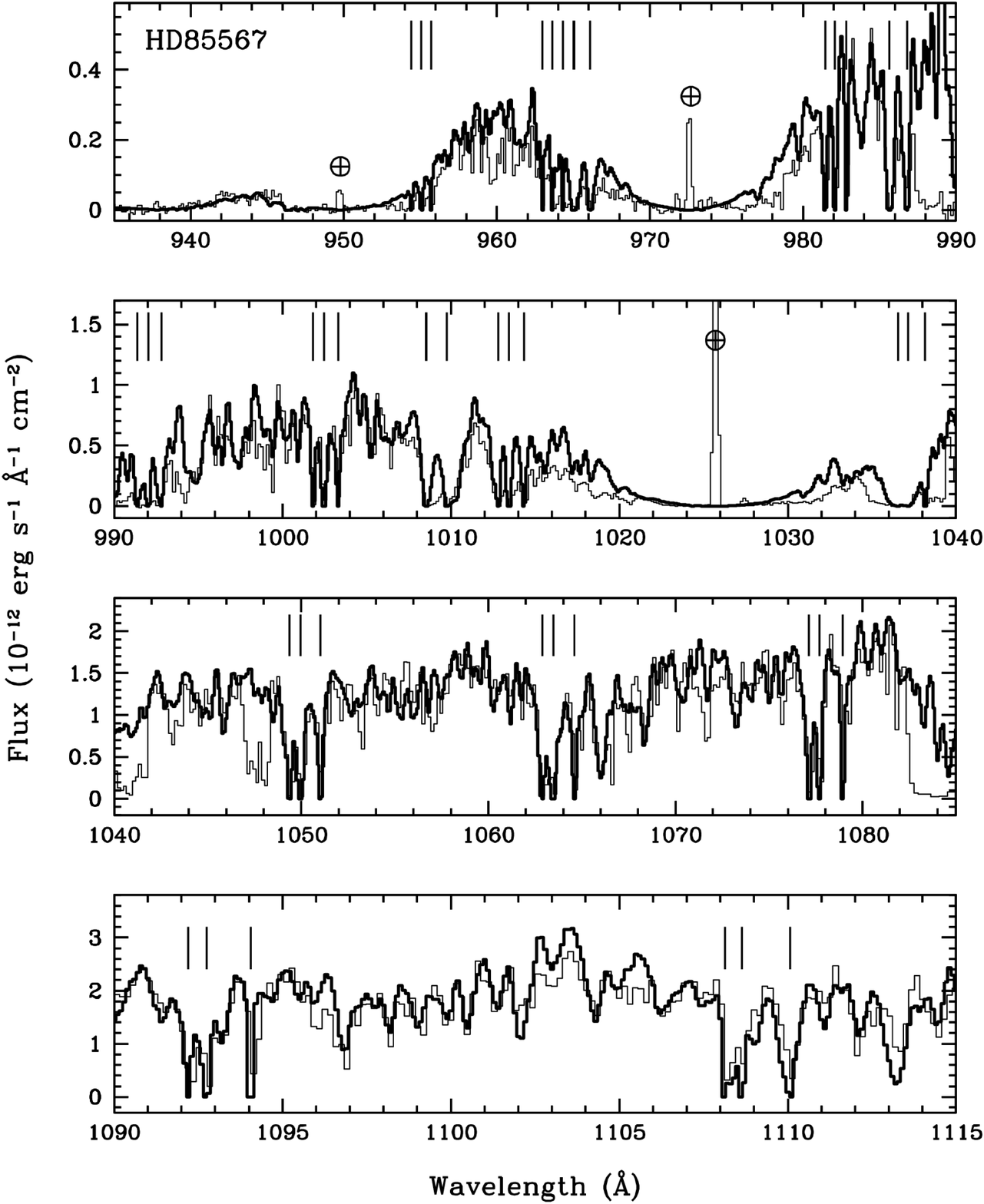}
\caption{Overplot of the \fuse\ spectrum of HD~85567 (thin line) and
  the synthetic spectrum obtained by combining our photospheric model
  with the results of the Meudon PDR Code (thick line). Same legend as
  for Fig.~\ref{spec_synt_hd176}.\vspace{2cm}}
      \label{spec_synt_hd85}
\end{figure*}
}

\onlfig{11}{
\begin{figure*}[!htbp] 
 \centering 
\vspace{2cm}
\includegraphics[width=16cm]{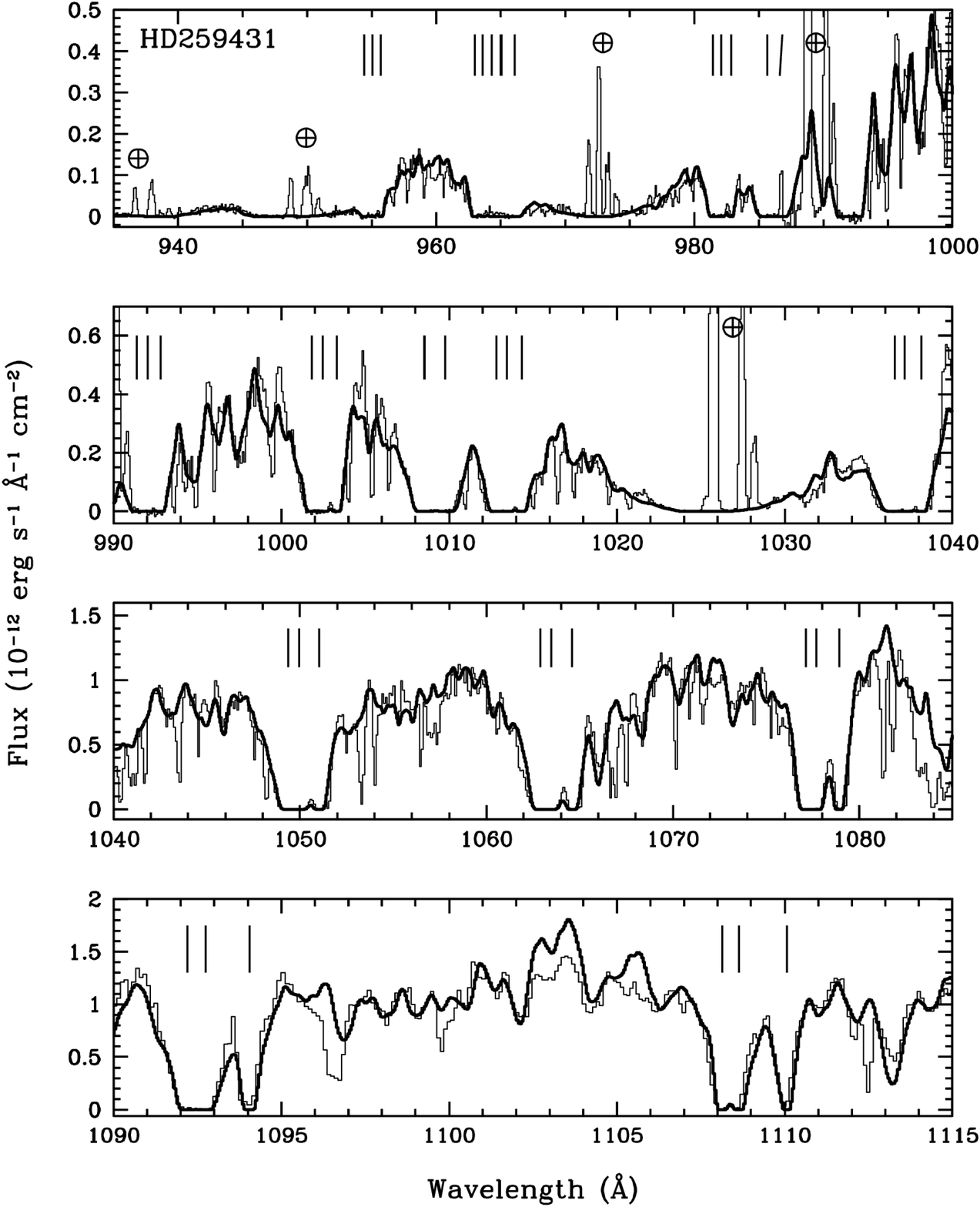}
\caption{Overplot of the \fuse\ spectrum of HD~259431 (thin line) and
  the synthetic spectrum obtained by combining our photospheric model
  with the results of the Meudon PDR Code (thick line). Same legend as
  for Fig.~\ref{spec_synt_hd176}.\vspace{2cm}}
      \label{spec_synt_hd259}
\end{figure*}
}

\onlfig{12}{
\begin{figure*}[!htbp] 
 \centering 
\vspace{2cm}
\includegraphics[width=16cm]{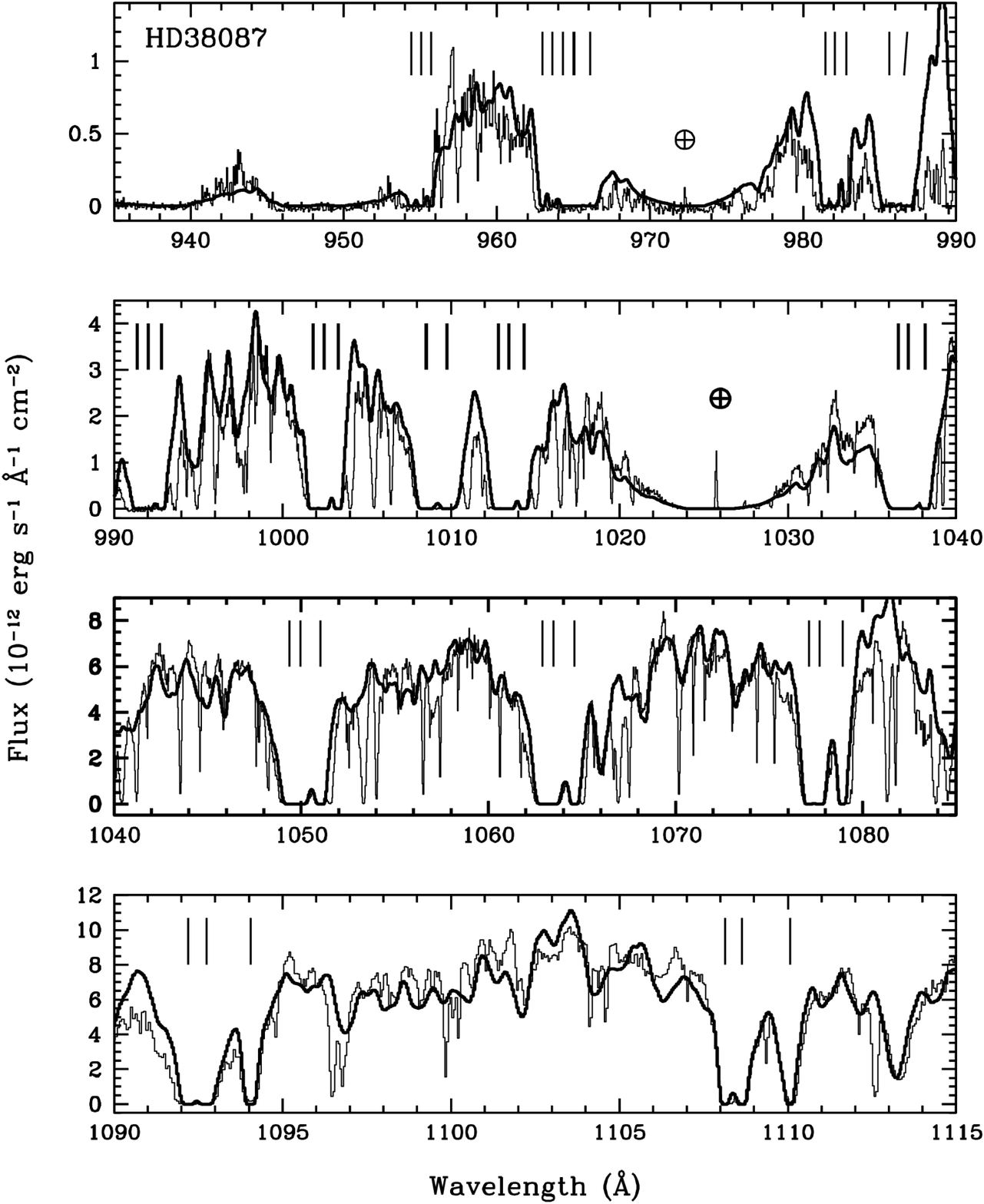}
\caption{Overplot of the \fuse\ spectrum of HD~38087 (thin line) and
  the synthetic spectrum obtained by combining our photospheric model
  with the results of the Meudon PDR Code (thick line). Same legend as
  for Fig.~\ref{spec_synt_hd176}. \vspace{2cm}}
      \label{spec_synt_hd38}
\end{figure*}
}


As shown in Fig.~\ref{PDR}, the general shape of the excitation
diagrams is reproduced well by PDR models with a single component, but
some details remain. The temperatures on the plots correspond to the
excitation temperatures of the low energy levels ($J=0-3$) given by
the best-fitting models. These temperatures are fully consistent with
those observed, and always fall within the error bars of the observed
excitation temperatures (Fig.~\ref{Be}). A relatively diffuse and cold
component reproduces the excitation of the lower J-levels, which
includes more than 90\%\ of the detected gas. However, we emphasize
that, as for the HAe stars, our modeling fails to reproduce perfectly
the excitation of the first J-levels of the ground vibrational level
and the higher J-levels ($J \geq 6$, $v=0$ and $J$, $v=1$)
simultaneously. By increasing the FUV field, one could expect to
increase the excitation of these levels, but this would affect the
column densities of the lower J-levels due to photodissociation. This
shows that UV pumping and formation pumping are probably not the only
mechanisms playing a role in the excitation of intermediate J-levels
we observe (see Sect.~\ref{H2}). However, at the present time, the
H$_2$ pumping formation process is not completely understood, and one
could expect that a slightly different modeling could vary the fit.

Using the photospheric models \citep[see Sect.~\ref{data} and
][]{JCB03} and the attenuation of the UV continuum provided by the PDR
code, we computed the synthetic spectrum of each star. To calculate
the spectrum, we only took the first three J-levels of H$_2$ into
account, which produce the strongest absorption lines.
Figure~\ref{spec_synt_hd76} shows the synthetic spectrum of HD~76534
overplotted on the observed \fuse\ spectrum. The synthetic spectra of
the other stars are available on-line (Fig.~\ref{spec_synt_hd176} to
Fig.~\ref{spec_synt_hd38}). The good agreement between the modeled and
the observed spectra strengthens confidence in our estimates of the
distance, density, molecular fraction, and gas-to-dust ratio we
estimated for each star from the different observational data.

\subsection{Additional information on the targets}

\begin{itemize}

\item {\bf HD~176386 -} This star is the primary star of a binary
  system \citep{Proust81, Vaz98}.  Our analysis of the FUV spectrum of
  the star is consistent with the star being surrounded by a large CS
  envelope {($d=0.04$~pc)}, a result in agreement with previous
  studies in different spectral domains. \citet{Siebenmorgen00} showed
  that the FWHM derived from the ISOCAM observations are indicative of
  a large extended halo around HD~176386, which was confirmed by the
  ISOPHOT multi-aperture sequence at 7.3$\mu m$.\\

\item {\bf HD~250550 -} In \cite{JCB03}, we showed that the properties
  of the H$_2$ and the atomic species derived from the \fuse\ spectrum
  of HD~250550 indicate that we are probing a dense CS environment,
  related to the remnant of the molecular cloud that collapsed to form
  the star. We refer the reader to this paper for more details.\\

\item {\bf HD~85567 -} Although our analysis does not allow us to rule
  out the presence of a CS disk, it does favor an interpretation in
  terms of a large CS envelope {($d=0.1$~pc)}. The presence of two
  components with different temperatures in the excitation diagram
  agrees with the \citet{Malfait98} study, which fit the SED of this
  star with a two-component, optically-thin, dusty envelope model. The
  apparent dip in the SED of the star between 6 and 10~$\mu m$ was
  interpreted as a physical hole in the dust distribution, caused by
  the break-up of an optically thin dusty disk \citep{Lada92}. In
  addition, \citet{MIRO01} have shown that, without imaging
  observations, it is impossible to distinguish between an extended,
  optically-thin, mostly spherical component, and a more compact
  disk. Those authors show that the emission-line profiles indicate
  complex structure in the CS envelope, which is likely non-spherical,
  with an optically-thick component and the possible presence of an
  extended optically-thin dusty component. We emphasize that if two
  components are present, we cannot distinguish them at the \fuse\
  resolution. \\
 
\item {\bf HD~259431 -} Our analysis, presented in a previous paper
  \citep{JCB03}, indicates that the material we observe is most likely
  related to a large CS envelope. This is consistent with the study by
  \citet{Malfait98}, who fit the SED of this star with a
  two-component, optically-thin, dusty envelope model. Although the
  mid-IR observations of HD~259431 by \cite{POLOMSKI02} are consistent
  with a model of a moderately flared CS disk, there is no clear
  observational evidence of a disk around this star.\\

\item {\bf HD~38087 -} \citet{Snow89} concluded from their analysis of
  the \iue\ spectrum that the star is surrounded by a dense envelope
  in which unusual grain growth has occurred. These results were
  confirmed by \citet{Burgh02}, who show that the region surrounding
  HD~38087 is dense and that the dust grains may be larger than
  average in the large nebula (NGC~2024) in which the star is
  located. From our modeling, we find a very low gas density and a
  large gas-to-star distance. The H$_2$ we observe in the \fuse\
  spectrum is likely located in the nebula NGC~2024, but the
  resolution of our observations does not allow us to rule out the
  presence of a second component corresponding to the star's
  envelope/disk.  \\

\item {\bf HD~76534 -} From a previous analysis \citep{klr04}, we
  concluded that, in the hostile CS environment of HD~76534, which is
  a B2 star, the presence of a CS disk is very unlikely. This agrees
  with the \citet{HILL92} analysis of the SED of HD~76534, who
  stressed the similarity of the very low NIR excess with that of a
  classical Be star, which is generally thought to be due to free-free
  emission in an ionized envelope rather than to CS dust
  \citep{HAMANN92_2}.

\end{itemize}

\section{Evolutionary trends}
\label{evol}

For each group of stars, we explored whether evolutionary trends could
be found in our results, especially between the total column densities
of H$_2$ in the circumstellar environment of the stars and the ages of
the stars.

\subsection{Ae/B9 stars}

\begin{figure}[!htbp] 
   \centering \includegraphics[width=9cm]{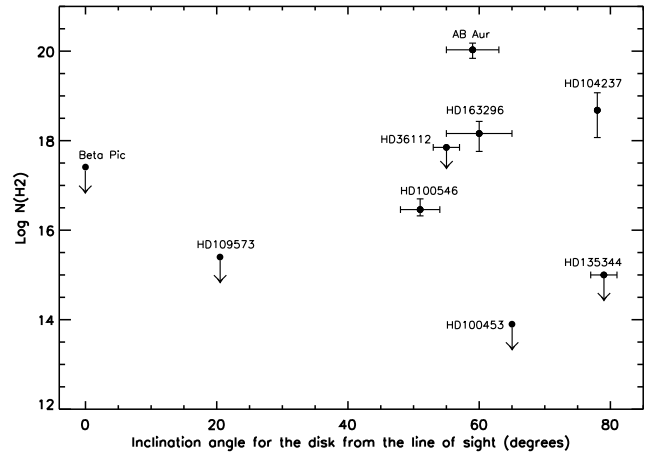}
   \caption{ Total H$_2$ column densities as a function of the
     inclination angles away from edge-on of the CS disks given in
     Table~\ref{results}. }
         \label{evol_disk}
\end{figure}

\begin{figure}[!htbp] 
   \centering \includegraphics[width=9cm]{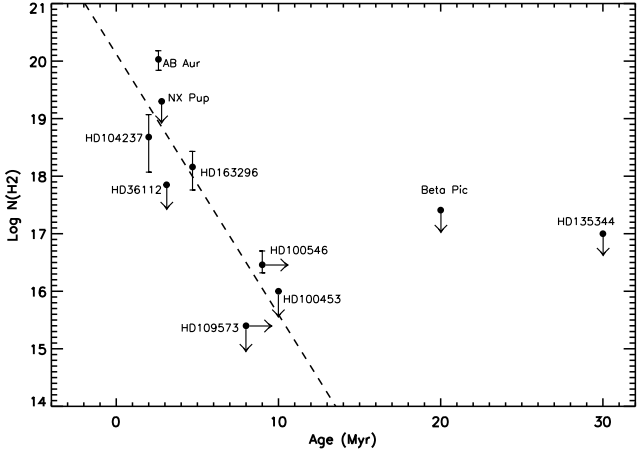}
   \caption{Total H$_2$ column densities as a function of the ages of
     those stars known to possess CS disks. Except for $\beta$
     Pictoris and HD~135344, for which we cannot conclude because we
     only have upper limits, the amount of H$_2$ is greater towards
     the youngest stars. The correlation between the age and the
     amount of gas is shown by the dashed line.}
         \label{evol_Ae}
\end{figure}

As we discussed in Sect.~\ref{disks}, the HAe stars form a very
inhomogeneous group. We plotted the column densities of H$_2$ towards
the HAe stars versus the inclination angles of their disks in
Fig.~\ref{evol_disk}, and in Fig.~\ref{evol_Ae} the column densities
of H$_2$ versus the ages of the HAes of our sample are presented.
Both Figs.~\ref{evol_disk} and \ref{evol_Ae} are consistent with CS
material that is not in disks. They are neither consistent with
non-associated interstellar clouds nor explained by absorption in the
disk material.

On the other hand, when excluding $\beta$ Pictoris and HD~135344, we
find that the amount of H$_2$ is higher for the younger stars than for
the older ones (Fig.~\ref{evol_Ae}). Such a correlation is very
unexpected if we observe non-CS media. However, our results do not
allow us to rule out CS envelopes around these stars. Indeed, for some
of these stars, the presence of both a disk and an envelope has been
demonstrated \citep[e.g.][]{GRADY01}. This would confirm the canonical
view of star formation, which predicts that the envelope becomes
progressively less important for obscuring the star and for dominating
the system mass with increasing age. The envelope dominates the mass
of the system in the early stages of stellar formation, but when the
star is close to the ZAMS, the envelope is expected to have dissipated
because of the action of stellar winds prior to central clearing of
the disk \citep[see reviews by][]{Calvet00, Mundy00}.

The correlation between the ages of the stars and the total amount of
H$_2$ needs to be confirmed. For most stars, we only estimated upper
limits on the column densities of H$_2$, and there are large
uncertainties on the ages of some of them . Only precise measurements
of the amount of H$_2$ and the ages of the stars could help to confirm
or deny the suspected correlation.

\subsection{Be stars}

\begin{figure}[!htbp] 
   \centering \includegraphics[width=9cm]{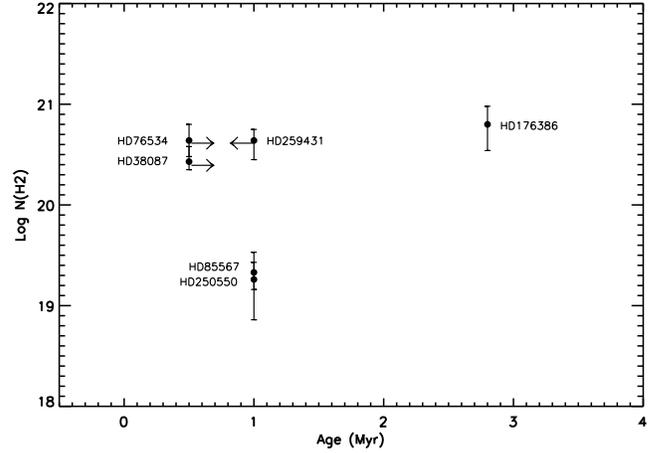}
   \caption{The total H$_2$ column densities as a function of the ages
     of the Be stars of our sample, which present no evidence of CS
     disks.}
         \label{evol_Be}
\end{figure}

Our analysis of the \fuse\ spectra of the Herbig Be stars shows that
this set of stars presents similar H$_2$ physical properties, favoring
an interpretation in terms of more or less spherical media around
these stars, very likely related to the original clouds in which the
stars formed. We find, however, no correlation between the total
column densities of H$_2$ and the age of the stars
(Fig.~\ref{evol_Be}), as one would expect from rapid dissipation of
the original envelope. This lack of correlation could be due to
different phenomena: (i) large contamination by the surrounding nebula
and thus the presence of different unresolved components along the
line of sight or (ii) uncertainty about the ages of the stars. Herbig
Be stars are more massive and more luminous than Ae stars, and the
time-scale to reach the ZAMS is very short, making their ages very
difficult to estimate.

\section{Summary and conclusions}
\label{discu}

We have distinguished two groups of stars: the stars known to possess
CS disks (Ae/B9 stars) and the stars with no evidence of disks (Be
stars). We interpreted the different excitation characteristics of
H$_2$ as direct probes of the origin of the observed gas: IS, CS
disks, or CS envelopes.

For most of the Herbig Ae/B9 stars, the inclination angles of the CS
disks from the lines of sight are quite high and not favorable to
probe the disks using absorption spectroscopy. Beta Pictoris and the
transitional object HD~109573 are exceptions. In the case of \bpic\,
we confirm the non-detection of H$_2$ in its \fuse\ spectrum given in
\cite{LECAV01}. We also did not detect H$_2$ in the spectrum of
HD~109573. These non-detections are consistent with the evolved status
of these systems and clearing of CS molecular gas from their disks.

AB Aurigae is clearly not edge-on, while H$_2$ is clearly detected,
but the cool component of the gas must be located at such a large
distance from the star that, even with a relatively large inclination
away from edge-on, some disk gas in the very outer regions of the disk
may still be in the line of sight.

In the sources HD~36112, HD~135344, and HD~100453, no H$_2$ is
observed. This is consistent with the high inclination angles of their
disks. The non-detections of gas imply that there are no detectable
remnant CS envelopes.

When molecular hydrogen gas is observed in the \fuse\ spectra of the
Herbig Ae/B9 stars of our sample, our analysis shows several kinds of
excitation, implying different origins and different physical
processes.

In one case, namely HD~141569, the origin of the detected gas is
clearly interstellar. For AB~Aur there is also some evidence of the
interstellar origin on the gas, but we lack the spectral resolution
that could allow us to probe the velocity distribution along the line
of sight and definitively decide between a CS or IS origin for the
gas. In spite of its relatively high inclination away from edge-on, it
is not geometrically impossible for the gas to originate in the upper
layers of the outer disk regions at about 800 AU, but the excitation
conditions of H$_2$ towards AB~Aur are fairly typical of what is
generally observed in the diffuse interstellar medium.  However, we
showed that the excitation diagram could not be reproduced well by a
single gaseous component along the line of sight. The large line width
value favors the presence of two components along the line of sight,
which are not resolved in our spectrum. In that case, one of these
components could correspond to the remnant of the molecular cloud in
which the star formed, as suggested by \cite{ROBERGE01}. The other
component should be hotter and probably closer to the star to explain
the excitation of the high J-levels of H$_2$.

For HD~100546, HD~163296, and HD~104237, we observed excited and
probably warm/hot, circumstellar H$_2$ that has excitation conditions
clearly different from those observed in the diffuse interstellar
medium. Such excitation conditions for H$_2$ give evidence of
collisionally excited media close to the stars. The high values of the
{\it b} parameters support this interpretation. In addition, the
measured radial velocities favor a CS origin for the H$_2$. However,
assuming that the gas and dust are coupled, the lines of sight towards
these three stars do not pass through their disks, and thus the H$_2$
we observed is not located in the disks. This raises questions about
the origin of the detected gas. One interesting possibility could be
that the H$_2$ gas is an FUV-driven photoevaporative wind from the
outer parts of the disk, but we are still missing clues needed for a
firm conclusion. The Meudon PDR code did not allow us to reproduce the
excitation of the observed H$_2$ gas. This implies that peculiar
excitation processes such as shocks or X-Rays, not taken into account
in the Meudon PDR Code, may play a role in the environment of these
stars.

The excitation conditions of circumstellar H$_2$ around the stars of
our second group (Be stars) are clearly different from those of the
first group. We found similar excitation conditions for the H$_2$ from
one star to the next. They present similarities to the conditions
found in the IS medium. Our analysis favors an interpretation in terms
of spherically symmetric media, not affected by inclination
effects. The excitation diagrams are reproduced nicely by PDR models,
at least for the cold component that includes more than 90\%\ of the
gas. On the other hand, these models do not simultaneously reproduce
the excitation conditions of the higher J-levels of H$_2$. We very
likely observe complex environments, such as large CS envelopes,
remnants of the original clouds in which the stars formed. However,
other excitation processes, in addition to those taken into account in
the Meudon PDR Code, are probably necessary to fully explain the
observed excitation diagrams.

Our FUV spectral analysis reinforces the differences between the two
subclasses of stars (Ae and Be) already highlighted by different
authors \citep[e.g.][]{NATTA00}. It also points to the need for
complementary observations to better understand the physical
conditions of formation and excitation of H$_2$ and the origin of the
gas. We suggest two avenues for bringing significant insight into
these issues. The first will be to observe spectral lines of the CH
and CH$^+$ molecules in the optical range at high spectral
resolution. These molecules are linked to the formation and excitation
of H$_2$ \citep{Federman82, Mattila86, Somerville89}. The formation of
CH is predicted to be controlled by gas-phase reactions with H$_2$.
The CH is thus a good tracer of H$_2$ and their abundances are
generally strongly correlated. In the presence of a shock, the
formation of the CH$^+$ molecule through the chemical reaction
C$^+$~+~H$_2$ needs a temperature about of 4500~K to occur. Thus, the
CH$^+$ molecule is a probe of hot and excited media, which could be
interpreted for our targets as material close to the star, and this
would allow us to better constrain the excitation of H$_2$. The
observation of CH and CH$^+$, if present, would provide direct
evidence of warm/hot CS gas close to the star. Combined with the H$_2$
data, this will allow us to explain the chemical mechanisms of
formation/destruction and excitation of the different molecules in the
CS environment of HAeBes and to better evaluate the velocity
dispersion along the line of sight (possibly due to turbulence).

The second avenue would be to observe pure rotational and
ro-vibrational transitions of H$_2$ in the infrared wavelength
range. This would give access to the spatial distribution and mass of
the observed gas and would help constrain the excitation
mechanisms. Such infrared observations would also provide information
about the circumstellar dust, including PAHs. Next-generation
instruments like VLT/VISIR could provide these kinds of observations
\citep{klr07a}.

All these data would give us the opportunity to obtain a global
picture of both the structure and the evolution of the CS environment
of HAeBes.


\begin{acknowledgements}
  This research is based on observations made with the NASA-CNES-CSA
  Far Ultraviolet Spectroscopic Explorer.  FUSE is operated for NASA
  by the Johns Hopkins University under NASA contract NAS5-32985.
  This work was done using the profile fitting procedure Owens.f
  developed by M. Lemoine and the \fuse\ French Team. C.M-Z. warmly
  thanks David Hollenbach for fruitful discussions about the possible
  sources of hot excited H$_2$. We thank D. Ehrenreich, B. Godard at
  JHU, and J.-C. Meunier at the LAM for their help in reprocessing the
  data.
\end{acknowledgements}


\bibliographystyle{aa}
\bibliography{Martin-Zaidi_fin}

\begin{thebibliography}{115}
\expandafter\ifx\csname natexlab\endcsname\relax\def\natexlab#1{#1}\fi

\bibitem[{{Abgrall} {et~al.}(2000){Abgrall}, {Roueff}, \& {Drira}}]{ABGRA3}
{Abgrall}, H., {Roueff}, E., \& {Drira}, I. 2000, \aaps, 141, 297

\bibitem[{{Abgrall} {et~al.}(1993{\natexlab{a}}){Abgrall}, {Roueff}, {Launay},
  {Roncin}, \& {Subtil}}]{ABGRA1}
{Abgrall}, H., {Roueff}, E., {Launay}, F., {Roncin}, J.~Y., \& {Subtil}, J.~L.
  1993{\natexlab{a}}, \aaps, 101, 273

\bibitem[{{Abgrall} {et~al.}(1993{\natexlab{b}}){Abgrall}, {Roueff}, {Launay},
  {Roncin}, \& {Subtil}}]{ABGRA2}
{Abgrall}, H., {Roueff}, E., {Launay}, F., {Roncin}, J.~Y., \& {Subtil}, J.~L.
  1993{\natexlab{b}}, \aaps, 101, 323

\bibitem[{{Acke} \& {Waelkens}(2004)}]{Acke04c}
{Acke}, B. \& {Waelkens}, C. 2004, \aap, 427, 1009

\bibitem[{{Augereau} {et~al.}(1999){Augereau}, {Lagrange}, {Mouillet}, \& {M{\'
  e}nard}}]{Augereau99}
{Augereau}, J.~C., {Lagrange}, A.~M., {Mouillet}, D., \& {M{\' e}nard}, F.
  1999, \aap, 350, L51

\bibitem[{{Augereau} {et~al.}(2001){Augereau}, {Lagrange}, {Mouillet}, \& {M{\'
  e}nard}}]{Augereau01}
{Augereau}, J.~C., {Lagrange}, A.~M., {Mouillet}, D., \& {M{\' e}nard}, F.
  2001, \aap, 365, 78

\bibitem[{{Augereau} \& {Papaloizou}(2004)}]{Augereau04}
{Augereau}, J.~C. \& {Papaloizou}, J.~C.~B. 2004, \aap, 414, 1153

\bibitem[{{Barrado y Navascu{\' e}s} {et~al.}(1999){Barrado y Navascu{\' e}s},
  {Stauffer}, {Song}, \& {Caillault}}]{Barrado99}
{Barrado y Navascu{\' e}s}, D., {Stauffer}, J.~R., {Song}, I., \& {Caillault},
  J.-P. 1999, \apjl, 520, L123

\bibitem[{{Bastien} \& {Menard}(1990)}]{Bastien90}
{Bastien}, P. \& {Menard}, F. 1990, \apj, 364, 232

\bibitem[{{Bertout} {et~al.}(1999){Bertout}, {Robichon}, \&
  {Arenou}}]{BERTOUT99}
{Bertout}, C., {Robichon}, N., \& {Arenou}, F. 1999, \aap, 352, 574

\bibitem[{{Beskrovnaya} {et~al.}(1999){Beskrovnaya}, {Pogodin},
  {Miroshnichenko}, {Th{\' e}}, {Savanov}, {Shakhovskoy}, {Rostopchina},
  {Kozlova}, \& {Kuratov}}]{Beskrovnaya99}
{Beskrovnaya}, N.~G., {Pogodin}, M.~A., {Miroshnichenko}, A.~S., {et~al.} 1999,
  \aap, 343, 163

\bibitem[{{Bitner} {et~al.}(2007){Bitner}, {Richter}, {Lacy}, {Greathouse},
  {Jaffe}, \& {Blake}}]{Bitner07}
{Bitner}, M.~A., {Richter}, M.~J., {Lacy}, J.~H., {et~al.} 2007, \apjl, 661,
  L69

\bibitem[{{Bohlin} {et~al.}(1978){Bohlin}, {Savage}, \& {Drake}}]{BOH78}
{Bohlin}, R.~C., {Savage}, B.~D., \& {Drake}, J.~F. 1978, \apj, 224, 132

\bibitem[{{B\"ohm} \& {Catala}(1993)}]{BC93}
{B\"ohm}, T. \& {Catala}, C. 1993, \aaps, 101, 629

\bibitem[{{B\"ohm} \& {Catala}(1995)}]{BC95}
{B\"ohm}, T. \& {Catala}, C. 1995, \aap, 301, 155

\bibitem[{{Boiss{\' e}} {et~al.}(2005){Boiss{\' e}}, {Le Petit}, {Rollinde},
  {Roueff}, {Pineau des For{\^ e}ts}, {Andersson}, {Gry}, \&
  {Felenbok}}]{Boisse05}
{Boiss{\' e}}, P., {Le Petit}, F., {Rollinde}, E., {et~al.} 2005, \aap, 429,
  509

\bibitem[{{Bouret} \& {Catala}(1998)}]{JCB98}
{Bouret}, J.-C. \& {Catala}, C. 1998, \aap, 340, 163

\bibitem[{{Bouret} {et~al.}(2002){Bouret}, {Deleuil}, {Lanz}, {Roberge},
  {Lecavelier des Etangs}, \& {Vidal-Madjar}}]{JCB02}
{Bouret}, J.-C., {Deleuil}, M., {Lanz}, T., {et~al.} 2002, \aap, 390, 1049

\bibitem[{{Bouret} {et~al.}(2003){Bouret}, {Martin}, {Deleuil}, {Simon}, \&
  {Catala}}]{JCB03}
{Bouret}, J.-C., {Martin}, C., {Deleuil}, M., {Simon}, T., \& {Catala}, C.
  2003, \aap, 410, 175

\bibitem[{{Bouwman} {et~al.}(2003){Bouwman}, {de Koter}, {Dominik}, \&
  {Waters}}]{Bouwman03}
{Bouwman}, J., {de Koter}, A., {Dominik}, C., \& {Waters}, L.~B.~F.~M. 2003,
  \aap, 401, 577

\bibitem[{{Browning} {et~al.}(2003){Browning}, {Tumlinson}, \&
  {Shull}}]{Browning03}
{Browning}, M.~K., {Tumlinson}, J., \& {Shull}, J.~M. 2003, \apj, 582, 810

\bibitem[{{Burgh} {et~al.}(2002){Burgh}, {McCandliss}, \& {Feldman}}]{Burgh02}
{Burgh}, E.~B., {McCandliss}, S.~R., \& {Feldman}, P.~D. 2002, \apj, 575, 240

\bibitem[{{Calvet} {et~al.}(2000){Calvet}, {Hartmann}, \& {Strom}}]{Calvet00}
{Calvet}, N., {Hartmann}, L., \& {Strom}, S.~E. 2000, Protostars and Planets
  IV, 377

\bibitem[{{Chen} \& {Kamp}(2004)}]{Chen04}
{Chen}, C.~H. \& {Kamp}, I. 2004, \apj, 602, 985

\bibitem[{{Chieze} {et~al.}(1998){Chieze}, {Pineau des Forets}, \&
  {Flower}}]{Chieze98}
{Chieze}, J.-P., {Pineau des Forets}, G., \& {Flower}, D.~R. 1998, \mnras, 295,
  672

\bibitem[{{Cidale} {et~al.}(2001){Cidale}, {Zorec}, \&
  {Tringaniello}}]{Cidale01}
{Cidale}, L., {Zorec}, J., \& {Tringaniello}, L. 2001, \aap, 368, 160

\bibitem[{{Corcoran} \& {Ray}(1997)}]{CORCORAN97}
{Corcoran}, M. \& {Ray}, T.~P. 1997, \aap, 321, 189

\bibitem[{{Crifo} {et~al.}(1997){Crifo}, {Vidal-Madjar}, {Lallement}, {Ferlet},
  \& {Gerbaldi}}]{crifo97}
{Crifo}, F., {Vidal-Madjar}, A., {Lallement}, R., {Ferlet}, R., \& {Gerbaldi},
  M. 1997, \aap, 320, L29

\bibitem[{{Deleuil} {et~al.}(2005){Deleuil}, {Bouret}, {Catala}, {Lecavelier
  des Etangs}, {Vidal-Madjar}, {Roberge}, {Feldman}, {Martin}, \&
  {Ferlet}}]{D05}
{Deleuil}, M., {Bouret}, J.-C., {Catala}, C., {et~al.} 2005, \aap, 429, 247

\bibitem[{{Deleuil} {et~al.}(2006){Deleuil}, {Bouret}, {Lecavelier Des Etangs},
  {Roberge}, \& {Feldman}}]{D06}
{Deleuil}, M., {Bouret}, J.~C., {Lecavelier Des Etangs}, A., {Roberge}, A., \&
  {Feldman}, P. 2006, in ASP Conf. Ser. 348: Astrophysics in the Far
  Ultraviolet: Five Years of Discovery with FUSE, ed. G.~{Sonneborn}, H.~W.
  {Moos}, \& B.-G. {Andersson}, 297--+

\bibitem[{{Deleuil} {et~al.}(2004){Deleuil}, {Lecavelier des Etangs}, {Bouret},
  {Roberge}, {Vidal-Madjar}, {Martin}, {Feldman}, \& {Ferlet}}]{D04}
{Deleuil}, M., {Lecavelier des Etangs}, A., {Bouret}, J.-C., {et~al.} 2004,
  \aap, 418, 577

\bibitem[{{Dent} {et~al.}(2005){Dent}, {Greaves}, \& {Coulson}}]{Dent05}
{Dent}, W.~R.~F., {Greaves}, J.~S., \& {Coulson}, I.~M. 2005, \mnras, 359, 663

\bibitem[{{Dixon} {et~al.}(2007){Dixon}, {Sahnow}, {Barrett}, {Civeit},
  {Dupuis}, {Fullerton}, {Godard}, {Hsu}, {Kaiser}, {Kruk}, {Lacour},
  {Lindler}, {Massa}, {Robinson}, {Romelfanger}, \& {Sonnentrucker}}]{Dixon07}
{Dixon}, W.~V., {Sahnow}, D.~J., {Barrett}, P.~E., {et~al.} 2007, \pasp, 119,
  527

\bibitem[{{Dominik} {et~al.}(2003){Dominik}, {Dullemond}, {Waters}, \&
  {Walch}}]{Dominik03}
{Dominik}, C., {Dullemond}, C.~P., {Waters}, L.~B.~F.~M., \& {Walch}, S. 2003,
  \aap, 398, 607

\bibitem[{{Donati} {et~al.}(1997){Donati}, {Semel}, {Carter}, {Rees}, \&
  {Collier Cameron}}]{Donati97}
{Donati}, J.-F., {Semel}, M., {Carter}, B.~D., {Rees}, D.~E., \& {Collier
  Cameron}, A. 1997, \mnras, 291, 658

\bibitem[{{Draine}(1978)}]{Draine78}
{Draine}, B.~T. 1978, \apjs, 36, 595

\bibitem[{{Dunkin} {et~al.}(1997){Dunkin}, {Barlow}, \& {Ryan}}]{Dunkin97}
{Dunkin}, S.~K., {Barlow}, M.~J., \& {Ryan}, S.~G. 1997, \mnras, 286, 604

\bibitem[{{Eisner} {et~al.}(2003){Eisner}, {Lane}, {Akeson}, {Hillenbrand}, \&
  {Sargent}}]{Eisner03}
{Eisner}, J.~A., {Lane}, B.~F., {Akeson}, R.~L., {Hillenbrand}, L.~A., \&
  {Sargent}, A.~I. 2003, \apj, 588, 360

\bibitem[{{Eisner} {et~al.}(2004){Eisner}, {Lane}, {Hillenbrand}, {Akeson}, \&
  {Sargent}}]{Eisner04}
{Eisner}, J.~A., {Lane}, B.~F., {Hillenbrand}, L.~A., {Akeson}, R.~L., \&
  {Sargent}, A.~I. 2004, \apj, 613, 1049

\bibitem[{{Falgarone} \& {Puget}(1995)}]{Falgarone95}
{Falgarone}, E. \& {Puget}, J.-L. 1995, \aap, 293, 840

\bibitem[{{Federman}(1982)}]{Federman82}
{Federman}, S.~R. 1982, \apj, 257, 125

\bibitem[{{Finkenzeller} \& {Jankovics}(1984)}]{FINK_JAN84}
{Finkenzeller}, U. \& {Jankovics}, I. 1984, \aaps, 57, 285

\bibitem[{{Gerbaldi} {et~al.}(1999){Gerbaldi}, {Faraggiana}, {Burnage},
  {Delmas}, {G{\' o}mez}, \& {Grenier}}]{Gerbaldi99}
{Gerbaldi}, M., {Faraggiana}, R., {Burnage}, R., {et~al.} 1999, \aaps, 137, 273

\bibitem[{{Grady} {et~al.}(2004){Grady}, {Woodgate B.}, {Torres C.A.O.},
  {Henning T.}, {APAI D.}, {RODMANN J.}, {WANG H.}, {STECKLUM B.}, {LINZ H.},
  {WILLIGER G.M.}, {BROWN A.}, {WILKINSON E.}, {HARPER G.M.}, {HERCZEG G.J.},
  {DANKS A.}, {VIEIRA G.L.}, {MALUMUTH E.}, \& {COLLINS N.R.~and HILL
  R.S.}}]{Grady04}
{Grady}, C., {Woodgate B.}, {Torres C.A.O.}, {et~al.} 2004, \apj, 608, 809

\bibitem[{{Grady} {et~al.}(2000){Grady}, {Devine}, {Woodgate}, {Kimble},
  {Bruhweiler}, {Boggess}, {Linsky}, {Plait}, {Clampin}, \& {Kalas}}]{GRADY00}
{Grady}, C.~A., {Devine}, D., {Woodgate}, B., {et~al.} 2000, \apj, 544, 895

\bibitem[{{Grady} {et~al.}(2001){Grady}, {Polomski}, {Henning}, {Stecklum},
  {Woodgate}, {Telesco}, {Pi{\~ n}a}, {Gull}, {Boggess}, {Bowers},
  {Bruhweiler}, {Clampin}, {Danks}, {Green}, {Heap}, {Hutchings}, {Jenkins},
  {Joseph}, {Kaiser}, {Kimble}, {Kraemer}, {Lindler}, {Linsky}, {Maran},
  {Moos}, {Plait}, {Roesler}, {Timothy}, \& {Weistrop}}]{GRADY01}
{Grady}, C.~A., {Polomski}, E.~F., {Henning}, T., {et~al.} 2001, \aj, 122, 3396

\bibitem[{{Gredel} {et~al.}(2002){Gredel}, {Pineau des For{\^ e}ts}, \&
  {Federman}}]{Gredel02}
{Gredel}, R., {Pineau des For{\^ e}ts}, G., \& {Federman}, S.~R. 2002, \aap,
  389, 993

\bibitem[{{Gry} {et~al.}(2002){Gry}, {Boulanger}, {Nehm{\' e}}, {Pineau des
  For{\^ e}ts}, {Habart}, \& {Falgarone}}]{Gry02}
{Gry}, C., {Boulanger}, F., {Nehm{\' e}}, C., {et~al.} 2002, \aap, 391, 675

\bibitem[{{H{\' e}brard} {et~al.}(2002){H{\' e}brard}, {Lemoine},
  {Vidal-Madjar}, {D{\' e}sert}, {Lecavelier des {\' E}tangs}, {Ferlet},
  {Wood}, {Linsky}, {Kruk}, {Chayer}, {Lacour}, {Blair}, {Friedman}, {Moos},
  {Sembach}, {Sonneborn}, {Oegerle}, \& {Jenkins}}]{HEBRARD02}
{H{\' e}brard}, G., {Lemoine}, M., {Vidal-Madjar}, A., {et~al.} 2002, \apjs,
  140, 103

\bibitem[{{Hamann} \& {Persson}(1992)}]{HAMANN92_2}
{Hamann}, F. \& {Persson}, S.~E. 1992, \apj, 394, 628

\bibitem[{{Herczeg} {et~al.}(2003){Herczeg}, {Linsky}, {Brown}, {Harper}, \&
  {Wilkinson}}]{Herczeg03}
{Herczeg}, G.~J., {Linsky}, J.~L., {Brown}, A., {Harper}, G.~M., \&
  {Wilkinson}, E. 2003, in The Future of Cool-Star Astrophysics: 12th Cambridge
  Workshop on Cool Stars, Stellar Systems, and the Sun (2001 July 30 - August
  3), eds. A. Brown, G.M. Harper, and T.R. Ayres, (University of Colorado),
  2003, p. 717-722., 717--722

\bibitem[{{Hern{\' a}ndez} {et~al.}(2004){Hern{\' a}ndez}, {Calvet}, {Brice{\~
  n}o}, {Hartmann}, \& {Berlind}}]{Hernandez04}
{Hern{\' a}ndez}, J., {Calvet}, N., {Brice{\~ n}o}, C., {Hartmann}, L., \&
  {Berlind}, P. 2004, \aj, 127, 1682

\bibitem[{{Hillenbrand} {et~al.}(1992){Hillenbrand}, {Strom}, {Vrba}, \&
  {Keene}}]{HILL92}
{Hillenbrand}, L.~A., {Strom}, S.~E., {Vrba}, F.~J., \& {Keene}, J. 1992, \apj,
  397, 613

\bibitem[{{Hollenbach} {et~al.}(1994){Hollenbach}, {Johnstone}, {Lizano}, \&
  {Shu}}]{Hollenbach94}
{Hollenbach}, D., {Johnstone}, D., {Lizano}, S., \& {Shu}, F. 1994, \apj, 428,
  654

\bibitem[{{Joulain} {et~al.}(1998){Joulain}, {Falgarone}, {Des Forets}, \&
  {Flower}}]{Joulain98}
{Joulain}, K., {Falgarone}, E., {Des Forets}, G.~P., \& {Flower}, D. 1998,
  \aap, 340, 241

\bibitem[{{Jura}(1991)}]{Jura91}
{Jura}, M. 1991, \apjl, 383, L79+

\bibitem[{{Juvela} {et~al.}(2002){Juvela}, {Mattila}, {Lehtinen}, {Lemke},
  {Laureijs}, \& {Prusti}}]{Juvela02}
{Juvela}, M., {Mattila}, K., {Lehtinen}, K., {et~al.} 2002, \aap, 382, 583

\bibitem[{{Lada}(1985)}]{Lada85}
{Lada}, C.~J. 1985, \araa, 23, 267

\bibitem[{{Lada} \& {Adams}(1992)}]{Lada92}
{Lada}, C.~J. \& {Adams}, F.~C. 1992, \apj, 393, 278

\bibitem[{{Lagage} {et~al.}(2006){Lagage}, {Doucet}, {Pantin}, {Habart},
  {Duch{\^e}ne}, {M{\'e}nard}, {Pinte}, {Charnoz}, \& {Pel}}]{Lagage06}
{Lagage}, P.-O., {Doucet}, C., {Pantin}, E., {et~al.} 2006, Science, 314, 621

\bibitem[{{Le Bourlot} {et~al.}(1999){Le Bourlot}, {Pineau des For{\^ e}ts}, \&
  {Flower}}]{LEBOURLOT99}
{Le Bourlot}, J., {Pineau des For{\^ e}ts}, G., \& {Flower}, D.~R. 1999,
  \mnras, 305, 802

\bibitem[{{Le Bourlot} {et~al.}(1993){Le Bourlot}, {Pineau Des Forets},
  {Roueff}, \& {Flower}}]{LEBOURLOT93}
{Le Bourlot}, J., {Pineau Des Forets}, G., {Roueff}, E., \& {Flower}, D.~R.
  1993, \aap, 267, 233

\bibitem[{{Le Petit} {et~al.}(2006){Le Petit}, {Nehm{\'e}}, {Le Bourlot}, \&
  {Roueff}}]{LePetit06}
{Le Petit}, F., {Nehm{\'e}}, C., {Le Bourlot}, J., \& {Roueff}, E. 2006, \apjs,
  164, 506

\bibitem[{{Lecavelier des Etangs} {et~al.}(2003){Lecavelier des Etangs},
  {Deleuil}, {Vidal-Madjar}, {Roberge}, {Le Petit}, {H{\' e}brard}, {Ferlet},
  {Feldman}, {D{\' e}sert}, \& {Bouret}}]{LECAV03}
{Lecavelier des Etangs}, A., {Deleuil}, M., {Vidal-Madjar}, A., {et~al.} 2003,
  \aap, 407, 935

\bibitem[{{Lecavelier des Etangs} {et~al.}(2001){Lecavelier des Etangs},
  {Vidal-Madjar}, {Roberge}, {Feldman}, {Deleuil}, {Andr{\' e}}, {Blair},
  {Bouret}, {D{\' e}sert}, {Ferlet}, {Friedman}, {H{\' e}brard}, {Lemoine}, \&
  {Moos}}]{LECAV01}
{Lecavelier des Etangs}, A., {Vidal-Madjar}, A., {Roberge}, A., {et~al.} 2001,
  \nat, 412, 706

\bibitem[{{Lehner} {et~al.}(2003){Lehner}, {Jenkins}, {Gry}, {Moos}, {Chayer},
  \& {Lacour}}]{Lehner03}
{Lehner}, N., {Jenkins}, E.~B., {Gry}, C., {et~al.} 2003, \apj, 595, 858

\bibitem[{{Leinert} {et~al.}(2001){Leinert}, {Haas}, {{\' A}brah{\' a}m}, \&
  {Richichi}}]{LEINERT01}
{Leinert}, C., {Haas}, M., {{\' A}brah{\' a}m}, P., \& {Richichi}, A. 2001,
  \aap, 375, 927

\bibitem[{{Lemoine} {et~al.}(2002){Lemoine}, {Vidal-Madjar}, {H{\' e}brard},
  {D{\' e}sert}, {Ferlet}, {Lecavelier des {\' E}tangs}, {Howk}, {Andr{\' e}},
  {Blair}, {Friedman}, {Kruk}, {Lacour}, {Moos}, {Sembach}, {Chayer},
  {Jenkins}, {Koester}, {Linsky}, {Wood}, {Oegerle}, {Sonneborn}, \&
  {York}}]{LEMOIN02}
{Lemoine}, M., {Vidal-Madjar}, A., {H{\' e}brard}, G., {et~al.} 2002, \apjs,
  140, 67

\bibitem[{{Malfait} {et~al.}(1998){Malfait}, {Bogaert}, \&
  {Waelkens}}]{Malfait98}
{Malfait}, K., {Bogaert}, E., \& {Waelkens}, C. 1998, \aap, 331, 211

\bibitem[{{Mannings} \& {Sargent}(1997)}]{Mannings97}
{Mannings}, V. \& {Sargent}, A.~I. 1997, \apj, 490, 792

\bibitem[{{Mannings} \& {Sargent}(2000)}]{Mannings00}
{Mannings}, V. \& {Sargent}, A.~I. 2000, \apj, 529, 391

\bibitem[{{Martin} {et~al.}(2004){Martin}, {Bouret}, {Deleuil}, {Simon}, \&
  {Catala}}]{klr04}
{Martin}, C., {Bouret}, J.-C., {Deleuil}, M., {Simon}, T., \& {Catala}, C.
  2004, \aap, 416, L5

\bibitem[{{Martin-Za{\"\i}di} {et~al.}(2005){Martin-Za{\"\i}di}, {Deleuil},
  {Simon}, {Bouret}, {Roberge}, {Feldman}, {Lecavelier Des Etangs}, \&
  {Vidal-Madjar}}]{klr05}
{Martin-Za{\"\i}di}, C., {Deleuil}, M., {Simon}, T., {et~al.} 2005, \aap, 440,
  921

\bibitem[{{Martin-Za{\"i}di} {et~al.}(2007){Martin-Za{\"i}di}, {Lagage},
  {Pantin}, \& {Habart}}]{klr07a}
{Martin-Za{\"i}di}, C., {Lagage}, P.-O., {Pantin}, E., \& {Habart}, E. 2007,
  \apjl, 666, L117

\bibitem[{{Mathis} {et~al.}(1977){Mathis}, {Rumpl}, \& {Nordsieck}}]{Mathis77}
{Mathis}, J.~S., {Rumpl}, W., \& {Nordsieck}, K.~H. 1977, \apj, 217, 425

\bibitem[{{Mattila}(1986)}]{Mattila86}
{Mattila}, K. 1986, \aap, 160, 157

\bibitem[{{Meeus} {et~al.}(2002){Meeus}, {Bouwman}, {Dominik}, {Waters}, \& {de
  Koter}}]{Meeus02}
{Meeus}, G., {Bouwman}, J., {Dominik}, C., {Waters}, L.~B.~F.~M., \& {de
  Koter}, A. 2002, \aap, 392, 1039

\bibitem[{{Meeus} {et~al.}(2001){Meeus}, {Waters}, {Bouwman}, {van den Ancker},
  {Waelkens}, \& {Malfait}}]{Meeus01}
{Meeus}, G., {Waters}, L.~B.~F.~M., {Bouwman}, J., {et~al.} 2001, \aap, 365,
  476

\bibitem[{{Millan-Gabet} {et~al.}(2001){Millan-Gabet}, {Schloerb}, \&
  {Traub}}]{Millan-Gabet01}
{Millan-Gabet}, R., {Schloerb}, F.~P., \& {Traub}, W.~A. 2001, \apj, 546, 358

\bibitem[{{Miroshnichenko} {et~al.}(2001){Miroshnichenko}, {Levato},
  {Bjorkman}, \& {Grosso}}]{MIRO01}
{Miroshnichenko}, A.~S., {Levato}, H., {Bjorkman}, K.~S., \& {Grosso}, M. 2001,
  \aap, 371, 600

\bibitem[{{Mora} {et~al.}(2001){Mora}, {Mer{\'{\i}}n}, {Solano}, {Montesinos},
  {de Winter}, {Eiroa}, {Ferlet}, {Grady}, {Davies}, {Miranda}, {Oudmaijer},
  {Palacios}, {Quirrenbach}, {Harris}, {Rauer}, {Cameron}, {Deeg}, {Garz{\'
  o}n}, {Penny}, {Schneider}, {Tsapras}, \& {Wesselius}}]{Mora01}
{Mora}, A., {Mer{\'{\i}}n}, B., {Solano}, E., {et~al.} 2001, \aap, 378, 116

\bibitem[{{Mundy} {et~al.}(2000){Mundy}, {Looney}, \& {Welch}}]{Mundy00}
{Mundy}, L.~G., {Looney}, L.~W., \& {Welch}, W.~J. 2000, Protostars and Planets
  IV, 355

\bibitem[{{Natta} {et~al.}(2000){Natta}, {Grinin}, \& {Mannings}}]{NATTA00}
{Natta}, A., {Grinin}, V., \& {Mannings}, V. 2000, Protostars and Planets IV,
  559

\bibitem[{{Palla} \& {Stahler}(1993)}]{PALLA93}
{Palla}, F. \& {Stahler}, S.~W. 1993, \apj, 418, 414

\bibitem[{{Pantin} {et~al.}(2005){Pantin}, {Bouwman}, \& {Lagage}}]{Pantin05}
{Pantin}, E., {Bouwman}, J., \& {Lagage}, P.~O. 2005, \aap, 437, 525

\bibitem[{{Pantin} {et~al.}(2000){Pantin}, {Waelkens}, \& {Lagage}}]{Pantin00}
{Pantin}, E., {Waelkens}, C., \& {Lagage}, P.~O. 2000, \aap, 361, L9

\bibitem[{{Polomski} {et~al.}(2002){Polomski}, {Telesco}, {Pi{\~ n}a}, \&
  {Schulz}}]{POLOMSKI02}
{Polomski}, E.~F., {Telesco}, C.~M., {Pi{\~ n}a}, R., \& {Schulz}, B. 2002,
  \aj, 124, 2207

\bibitem[{{Proust} {et~al.}(1981){Proust}, {Ochsenbein}, \&
  {Pettersen}}]{Proust81}
{Proust}, D., {Ochsenbein}, F., \& {Pettersen}, B.~R. 1981, \aaps, 44, 179

\bibitem[{{Redfield} \& {Linsky}(2004)}]{Redfield04}
{Redfield}, S. \& {Linsky}, J.~L. 2004, \apj, 613, 1004

\bibitem[{{Rice} {et~al.}(2003){Rice}, {Wood}, {Armitage}, {Whitney}, \&
  {Bjorkman}}]{Rice03}
{Rice}, W.~K.~M., {Wood}, K., {Armitage}, P.~J., {Whitney}, B.~A., \&
  {Bjorkman}, J.~E. 2003, \mnras, 342, 79

\bibitem[{{Richter} {et~al.}(2002){Richter}, {Jaffe}, {Blake}, \&
  {Lacy}}]{Richter02}
{Richter}, M.~J., {Jaffe}, D.~T., {Blake}, G.~A., \& {Lacy}, J.~H. 2002, \apjl,
  572, L161

\bibitem[{{Roberge} {et~al.}(2006){Roberge}, {Feldman}, {Weinberger},
  {Deleuil}, \& {Bouret}}]{Roberge06}
{Roberge}, A., {Feldman}, P.~D., {Weinberger}, A.~J., {Deleuil}, M., \&
  {Bouret}, J.-C. 2006, \nat, 441, 724

\bibitem[{{Roberge} {et~al.}(2001){Roberge}, {Lecavelier des Etangs}, {Grady},
  {Vidal-Madjar}, {Bouret}, {Feldman}, {Deleuil}, {Andre}, {Boggess},
  {Bruhweiler}, {Ferlet}, \& {Woodgate}}]{ROBERGE01}
{Roberge}, A., {Lecavelier des Etangs}, A., {Grady}, C.~A., {et~al.} 2001,
  \apjl, 551, L97

\bibitem[{{Royer} {et~al.}(2002){Royer}, {Grenier}, {Baylac}, {G{\' o}mez}, \&
  {Zorec}}]{Royer02b}
{Royer}, F., {Grenier}, S., {Baylac}, M.-O., {G{\' o}mez}, A.~E., \& {Zorec},
  J. 2002, \aap, 393, 897

\bibitem[{{Sako} {et~al.}(2005){Sako}, {Yamashita}, {Kataza}, {Miyata},
  {Okamoto}, {Honda}, {Fujiyoshi}, \& {Onaka}}]{Sako05}
{Sako}, S., {Yamashita}, T., {Kataza}, H., {et~al.} 2005, \apj, 620, 347

\bibitem[{{Semenov} {et~al.}(2005){Semenov}, {Pavlyuchenkov}, {Schreyer},
  {Henning}, {Dullemond}, \& {Bacmann}}]{Semenov05}
{Semenov}, D., {Pavlyuchenkov}, Y., {Schreyer}, K., {et~al.} 2005, \apj, 621,
  853

\bibitem[{{Setiawan} {et~al.}(2008){Setiawan}, {Henning}, {Launhardt},
  {M{\"u}ller}, {Weise}, \& {K{\"u}rster}}]{Setiawan08}
{Setiawan}, J., {Henning}, T., {Launhardt}, R., {et~al.} 2008, \nat, 451, 38

\bibitem[{{Shull} \& {Beckwith}(1982)}]{Shull82}
{Shull}, J.~M. \& {Beckwith}, S. 1982, \araa, 20, 163

\bibitem[{{Siebenmorgen} {et~al.}(2000){Siebenmorgen}, {Prusti}, {Natta}, \&
  {M{\" u}ller}}]{Siebenmorgen00}
{Siebenmorgen}, R., {Prusti}, T., {Natta}, A., \& {M{\" u}ller}, T.~G. 2000,
  \aap, 361, 258

\bibitem[{{Smith} \& {Terrile}(1984)}]{Smith84}
{Smith}, B.~A. \& {Terrile}, R.~J. 1984, Science, 226, 1421

\bibitem[{{Snow} \& {McCall}(2006)}]{Snow06}
{Snow}, T.~P. \& {McCall}, B.~J. 2006, \araa, 44, 367

\bibitem[{{Snow} \& {Witt}(1989)}]{Snow89}
{Snow}, T.~P. \& {Witt}, A. 1989, \apj, 342, 295

\bibitem[{{Somerville} \& {Smith}(1989)}]{Somerville89}
{Somerville}, W.~B. \& {Smith}, C.~A. 1989, \mnras, 238, 559

\bibitem[{{Spitzer} \& {Cochran}(1973)}]{Spitzer73}
{Spitzer}, L.~J. \& {Cochran}, W.~D. 1973, \apjl, 186, L23+

\bibitem[{{Stapelfeldt} {et~al.}(1998){Stapelfeldt}, {Burrows}, {Krist},
  {Watson}, {Ballester}, {Clarke}, {Crisp}, {Evans}, {Gallagher}, {Griffiths},
  {Hester}, {Hoessel}, {Holtzman}, {Mould}, {Scowen}, {Trauger}, \&
  {Westphal}}]{Stapelfeldt98}
{Stapelfeldt}, K.~R., {Burrows}, C.~J., {Krist}, J.~E., {et~al.} 1998, \apj,
  508, 736

\bibitem[{{Telesco} {et~al.}(2000){Telesco}, {Fisher}, {Pi{\~ n}a}, {Knacke},
  {Dermott}, {Wyatt}, {Grogan}, {Holmes}, {Ghez}, {Prato}, {Hartmann}, \&
  {Jayawardhana}}]{Telesco00}
{Telesco}, C.~M., {Fisher}, R.~S., {Pi{\~ n}a}, R.~K., {et~al.} 2000, \apj,
  530, 329

\bibitem[{{Testi} {et~al.}(1998){Testi}, {Palla}, \& {Natta}}]{TESTI98}
{Testi}, L., {Palla}, F., \& {Natta}, A. 1998, \aaps, 133, 81

\bibitem[{{Th\'e} {et~al.}(1994){Th\'e}, {de Winter}, \& {Perez}}]{THE94}
{Th\'e}, P.~S., {de Winter}, D., \& {Perez}, M.~R. 1994, \aaps, 104, 315

\bibitem[{{Thi} {et~al.}(2001){Thi}, {van Dishoeck}, {Blake}, {van Zadelhoff},
  {Horn}, {Becklin}, {Mannings}, {Sargent}, {van den Ancker}, {Natta}, \&
  {Kessler}}]{Thi01}
{Thi}, W.~F., {van Dishoeck}, E.~F., {Blake}, G.~A., {et~al.} 2001, \apj, 561,
  1074

\bibitem[{{Torres} {et~al.}(2003){Torres}, {Guenther}, {Marschall}, {Neuh{\"
  a}user}, {Latham}, \& {Stefanik}}]{Torres03}
{Torres}, G., {Guenther}, E.~W., {Marschall}, L.~A., {et~al.} 2003, \aj, 125,
  825

\bibitem[{{Valenti} {et~al.}(2000){Valenti}, {Johns-Krull}, \&
  {Linsky}}]{valenti00}
{Valenti}, J.~A., {Johns-Krull}, C.~M., \& {Linsky}, J.~L. 2000, \apjs, 129,
  399

\bibitem[{{van den Ancker} {et~al.}(1998){van den Ancker}, {de Winter}, \&
  {Tjin A Djie}}]{VdA98}
{van den Ancker}, M.~E., {de Winter}, D., \& {Tjin A Djie}, H.~R.~E. 1998,
  \aap, 330, 145

\bibitem[{{Vaz} {et~al.}(1998){Vaz}, {Andersen}, {Casey}, {Clausen}, {Mathieu},
  \& {Heyer}}]{Vaz98}
{Vaz}, L.~P.~R., {Andersen}, J., {Casey}, B.~W., {et~al.} 1998, \aaps, 130, 245

\bibitem[{{Wahhaj} {et~al.}(2005){Wahhaj}, {Koerner}, {Backman}, {Werner},
  {Serabyn}, {Ressler}, \& {Lis}}]{Wahhaj05}
{Wahhaj}, Z., {Koerner}, D.~W., {Backman}, D.~E., {et~al.} 2005, \apj, 618, 385

\bibitem[{{Weinberger} {et~al.}(1999){Weinberger}, {Becklin}, {Schneider},
  {Smith}, {Lowrance}, {Silverstone}, {Zuckerman}, \& {Terrile}}]{Weinberger99}
{Weinberger}, A.~J., {Becklin}, E.~E., {Schneider}, G., {et~al.} 1999, \apjl,
  525, L53

\end{thebibliography}



\end{document}